\documentclass[%
 reprint,
superscriptaddress,
 amsmath,amssymb,
 aps,
]{revtex4-2}

\usepackage{bbold}

\usepackage{comment}
\usepackage{graphicx}
\usepackage{dcolumn}
\usepackage{bm}
\usepackage{color}
\usepackage[caption=false]{subfig}

\usepackage{xcolor}
\usepackage[colorlinks=true,linkcolor=blue,citecolor=blue,urlcolor=blue]{hyperref}

\begin{document}

\preprint{APS/123-QED}

\title{Highly robust logical qubit encoding in an ensemble of V-symmetrical qutrits}


\author{Luis Octavio Casta\~{n}os-Cervantes}
\affiliation{Tecnologico de Monterrey, School of Engineering and Sciences, 14380, Mexico City, Mexico}%
\email{luis.castanos@tec.mx}
\affiliation{Universidad Nacional Aut\'{o}noma de M\'{e}xico, Facultad de Ingenier\'{i}a, Av. Universidad 3000, Ciudad Universitaria, Coyoac\'{a}n 04510, Ciudad de M\'{e}xico, M\'{e}xico}
\author{Manuel Calixto}
\affiliation{Department of Applied Mathematics, University of Granada, Fuentenueva s/n, 18071 Granada, Spain}%
\email{calixto@ugr.es}
\affiliation{Institute Carlos I for Theoretical and Computational Physics (IC1), Fuentenueva s/n, 18071 Granada, Spain}%
\author{Julio Guerrero} 
\affiliation{Department of Mathematics, University of Ja\'{e}n, Campus Las Lagunillas s/n, 23071, Spain}%
\email{jguerrer@ujaen.es}
\affiliation{Institute Carlos I for Theoretical and Computational Physics (IC1), Fuentenueva s/n, 18071 Granada, Spain}%

%
%

\date{\today}

\begin{abstract}
We propose using even and odd Sch\"{o}dinger cat states formed from coherent states of U(3) of an ensemble of qutrits with a symmetrical V-configuration (\textit{a qubit-disguised qutrit}) to encode a logical qubit. These carefully engineered logical qubit states are parameter independent stationary states of the effective master equation governing the evolution of the ensemble and, consequently, constitute dark states and are invulnerable to dissipation and correlated collective dephasing. In particular, the logical qubit states are immune to single qutrit decay (the analogous of single photon loss process for qutrits) and simultaneous decay and driving of two qutrits (the analogous two-photon loss and driving processes for qutrits). In addition, we show how to implement the single-qubit quantum NOT gate and the Hadamard gate followed by either the phase gate or the phase and $Z$ gates. We study analytically the case of two qutrits and conclude that the logical qubit states exhibit parity-sensitive inhomogeneous broadening and local correlated dephasing: the even logical state is completely immune to these processes, while odd one is vulnerable. Nevertheless, in the presence of these interactions one can also define another odd state with mixed permutation symmetry that is immune to both inhomogeneous broadening and local correlated dephasing. We suggest that these results can be extrapolated to an arbitrary number of qutrits. The effective master equation is deduced from a physical system composed of two parametrically coupled cavities with one of them interacting dispersively with an ensemble of three-level atoms (the qutrits). A proof that the logical qubit is also part of a stationary solution of the physical system under the considered approximations is given. In principle this physical system can be implemented by means of two coplanar waveguide resonators, a SQUID parametrically coupling them, and a cloud of alkali atoms close to one of the resonators.
\end{abstract}

\maketitle


\section{Introduction}

In quantum information theory it is common to restrict the state space of a physical system to a two dimensional subspace in order to represent a qubit \cite{Chuang}. Alternatively, one can use a physical system with a higher dimensional state space and encode the logical states of a qubit as two orthogonal pure states \cite{Cochrane,PowerCats,Mirrahimi}. This is especially attractive because one can then design a physical system where the encoded qubit logical states can be robust against dissipation and dephasing \cite{Mirrahimi,Mirrahimi2015,RydbergRMP,RydbergReview,RydbergQubit,NoriO,squeezing,Nori2024,IHBnori}. A paradigmatic example consists in encoding the logical states of a qubit in a pair of even and odd  (with respect to a suitable parity operator) Schr\"{o}dinger cat states of a harmonic oscillator \cite{Cochrane,Mirrahimi}. In the case of a single-mode electromagnetic field, these cat states can be protected against dephasing by means of two-photon loss processes but they are vulnerable to single photon loss and this can limit their use in applications \cite{Mirrahimi,Mirrahimi2015}.

Another attractive possibility is to encode a qubit in an ensemble of atoms or spins because this type of systems can have long coherence times \cite{RydbergRMP,RydbergReview,RydbergQubit,NoriO,squeezing,Nori2024}. Some of their major sources of noise are collective and local dephasing (that is, population-conserving scattering processes) and inhomogeneous broadening. For example, inhomogeneous broadening due to the Doppler effect and dephasing caused by collisions and laser fluctuations can obstruct the observation of  coherent phenomena in atomic gases \cite{EITreview,Dephasing1,DephasingJCS}. Some promising approaches rely on the use of Rydberg atoms \cite{RydbergRMP,RydbergReview,RydbergQubit}, while other implementations consider a hybrid system of superconducting circuits and spin ensembles of either alkali atoms \cite{squeezing,RydbergQubit} or nitrogen-vacancy centers \cite{squeezing,Nori2024}. The former are based on the Rydberg dipole blockade mechanism and on multiphoton processes \cite{RydbergRMP,RydbergReview,RydbergBlockade}. Moreover, for these systems it has been demonstrated that an encoded qubit can be controlled coherently and that it can be robust both to depletion of atoms and to noise \cite{RydbergQubit}.

Especially relevant to this article is the hybrid system approach, since \cite{NoriO} proposed a two-qubit simultaneous decay process that conserves the parity of the number of the excited qubits and that can stabilize Schr\"{o}dinger cat states of spin coherent states in the ensemble. In order to achieve this mechanism analogous to the two-photon loss process, Ref.~\cite{NoriO} considered a system of two parametrically coupled single-mode cavities, one containing a driven \textit{pump field} and the other containing a \textit{signal field} that interacts dispersively with an ensemble of qubits. The effective master equation describing the evolution of the ensemble contains the desired two-qubit simultaneous decay process, as well as a linear dissipator which describes single qubit decay, a process analogous to single photon loss. Assuming that the number of excited qubits is much smaller than the total number of qubits and fine tuning the system parameters, Ref.~\cite{NoriO} found that the linear dissipator can be neglected and that the qubit ensemble is driven to an even (odd) cat state if the ensemble is initially in an even (odd) parity state. Quite remarkably, this leads to qubit Schr\"{o}dinger cat states that can have a lifetime several orders of magnitude longer than cavity photonic cat states. By considering the same system but without driving and without the linear dissipator, Ref.~\cite{Nori2024} then showed that the system relaxes with high fidelity to states that can also be used as qubit logical states. In addition, it was also found \cite{IHBnori} that the effects of inhomogeneous broadening depends on the parity and amplitude of the cat states, with the even cat state being significantly more robust against dephasing than the odd one for small amplitudes.

The same system presented in \cite{NoriO} has also been used to propose a protocol for the generation of spin squeezing with and without driving the pump field and obtained results comparable to those of the two-axis twisting model \cite{squeezing}. Furthermore, Ref.~\cite{squeezing} discussed in detail the experimental feasibility of the protocol and proposed two implementations using two superconducting coplanar waveguide resonators (CPWR), a superconducting quantum interference device (SQUID), and an ensemble of either rubidium 87 atoms or nitrogen-vacancy  centers in diamond. The SQUID is in charge of the parametric coupling between the cavities and the ensemble is placed above one of them and is magnetically coupled to it. In particular, \cite{Rb87CPWR} has already demonstrated both the dispersive and resonant magnetic coupling of an ensemble of ultracold rubidium 87 atoms to a CPWR.  

In this article we consider the same system as that introduced in \cite{NoriO} but with an ensemble of $N \geq 2$ qutrits with a V-type configuration (\textit{a qubit-disguised qutrit}) replacing the qubits. The objective is to investigate if the added degree of freedom can be used to protect the logical qubit states from single qutrit decay, that is, from the process analogous to single-photon loss. The results are quite remarkable. If one chooses the logical qubit states as specific even and odd Schr\"{o}dinger cat states of U(3) with a definite parity of the number of particles in the ground level, then it turns out that these logical states are parameter independent stationary states of the effective master equation that describes the evolution of the ensemble of qutrits. Consequently, they are dark states and are immune to single qutrit decay and to correlated collective dephasing. In addition, analytical results for two qutrits show that they exhibit parity-sensitive inhomogeneous broadening and local correlated dephasing: the even cat state is immune to both, while the odd cat state is vulnerable. In the presence of these permutation symmetry breaking interactions it turns out that one can consider another mixed permutation symmetry logical state which is immune to these processes. Unfortunately, the trade off is that the system does not always relax to the logical states. However, they are easy to create with a convenient driving process. In principle, the system considered in this article could be implemented experimentally by using the same physical system proposed in \cite{squeezing}, since the CPWR used in \cite{Rb87CPWR} naturally couples three ground state hyperfine levels of rubidium 87 in a V-configuration. The problem is that the coupling strength is weak and would have to be increased, for example, by decreasing the distance between the CPWR and the ensemble of atoms \cite{Rb87CPWR} or by using a state of the art CPWR \cite{CPWRnew}.

The article is structured as follows. In Sec.~II we introduce the system under study and establish the effective master equation that describes its evolution. In Sec.~III we present the parity symmetry of the master equation. Then, in Sec.~IV we characterize the parameter independent stationary states of the master equation and use them to define the logical qubit states. In Sec.~V we describe how to implement a quantum NOT gate and a Hadamard gate followed by a phase gate or a phase and $\mathcal{Z}$ gates and in Sec.~VI we consider the case of two qutrits and determine the effects of local dephasing and inhomogeneous broadening. In Sec.~VII we deduce the effective master equation of Sec.~II from a physical system composed of two parametrically coupled cavities and an ensemble of three-level atoms. Finally, the conclusions are in Sec.~VIII.

\section{The model}
\label{SecQutrits}

\begin{figure}[b]
\includegraphics[scale=0.9]{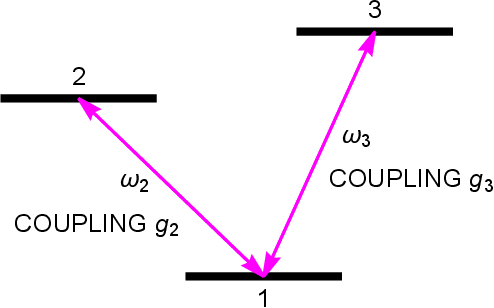}
\caption{\label{Figure1} Each qutrit is a quantum $3$-level system with a V-configuration where $\vert 1\rangle$ is the ground level and $\vert 2 \rangle$ and $\vert 3 \rangle$ are the excited levels. The angular frequency associated with the transition $\vert j \rangle \leftrightarrow \vert 1 \rangle$ is $\omega_{j} >0$ with $j=2,3$. In the context of the physical system of Sec.~\ref{Origen}, the qutrits are coupled to a single-mode cavity quantum electromagnetic field called the \textit{signal field}. The coupling strengths are $g_{2}$ and $g_{3}$.}
\end{figure}

Consider a system consisting of $N \geq 2$ identical qutrits. Each qutrit is a quantum 3-level system with a V-configuration where $\vert 1 \rangle$  is the ground level and $\vert 2 \rangle$ and $\vert 3 \rangle$ are the excited levels, see Fig.~\ref{Figure1}. The angular transition frequency between levels $\vert j \rangle$ and $\vert 1 \rangle$ is denoted by $\omega_{j}$, whereas direct transitions between levels $\vert 2 \rangle$ and $\vert 3 \rangle$ are forbidden. Assume that the qutrits are bosons and that $\omega_{3} = \omega_{2} = \omega_{q} >0$, that is,  the qutrits have a symmetrical V-configuration. 

The ensemble of qutrits is described in second quantization where $b_{j}^{\dagger}$ ($b_{j}$) creates (annihilates) a particle in the level $\vert j \rangle$ $(j=1,2,3)$. These operators satisfy the commutation relations $[b_{j},b_{k}] = 0$ and $[b_{j},b_{k}^{\dagger}] = \delta_{kj}$ for $j,k=1,2,3$.  Here $\delta_{kj}$ is the Kronecker delta.

An orthonormal basis $\beta_{q}$ for the state space of the qutrits is obtained by using the occupation number states $\vert n_{1},n_{2},n_{3} \rangle$ where there are $n_{j}$ qutrits in the level $\vert j \rangle$ $(j=1,2,3)$:
\begin{eqnarray}
\label{betaa}
\beta_{q} &=& \left\{ \vert n_{1},n_{2},n_{3} \rangle: \ n_{1},n_{2},n_{3}\in\mathbb{Z}^{+}, \ \sum_{j=1}^{3} n_{j} = N \right\} , \quad
\end{eqnarray}
with $\mathbb{Z}^{+}$ the set of nonnegative integers. Then, one has
\begin{eqnarray}
\label{bj}
b_{j}^{\dagger} \vert n_{1},n_{2},n_{3} \rangle &=& \sqrt{n_{j}+1} \vert n_{1}+\delta_{j1}, n_{2} +\delta_{j2}, n_{3}+\delta_{j3} \rangle , \cr
b_{j} \vert n_{1},n_{2},n_{3} \rangle &=& \sqrt{n_{j}} \vert n_{1}-\delta_{j1}, n_{2} -\delta_{j2}, n_{3}-\delta_{j3} \rangle ,
\end{eqnarray}
for $j=1,2,3$.

For $j,k = 1,2,3$ we introduce the operators
\begin{eqnarray}
\label{Sij}
S_{jk} = b_{j}^{\dagger}b_{k}, \quad S_{-} = S_{12} + S_{13} .
\end{eqnarray}
Notice that $S_{jk}^{\dagger} = S_{kj}$, $S_{jj}$ counts the number of qutrits in the level $\vert j \rangle$, and $S_{jk}$ with $k\not= j$ annihilates a qutrit in the level $\vert k \rangle$ and creates one in the level $\vert j \rangle$. Also, $S_{-}$ applied to an occupation number state $\vert n_{1},n_{2},n_{3} \rangle$ gives rise to a superposition of two occupation number states, one where a qutrit made a transition from the level $\vert 2 \rangle$ to the level $\vert 1 \rangle$ plus another where a qutrit made a transition from the level $\vert 3 \rangle$ to the level $\vert 1 \rangle$. In this sense $S_{-}$ can be thought of as an operator that describes the transition of one qutrit from an excited level to the ground level.

The effective master equation describing the evolution of the ensemble of qutrits is 
\begin{eqnarray}
\label{EcMaestraFinal}
\frac{d}{dt} \rho (t) 
&=& \mathcal{L} \rho (t) ,
\end{eqnarray}
where $\rho (t)$ is the density operator of the ensemble and the Liouvillian $\mathcal{L}$ is defined by
\begin{eqnarray}
\label{EcMaestraFinalb}
\mathcal{L}A
&=& -\frac{i}{\hbar} \left[ H_{q}, \ A \right] + \kappa_{1} \mathcal{D}(S_{-}) A + \kappa_{2} \mathcal{D}(S_{-}^{2}) A .
\end{eqnarray}
Here $\kappa_{1}$, $\kappa_{2} > 0$ and the superoperator $\mathcal{D}(\cdot)$ is defined by
 \begin{eqnarray}
 \label{EcMaestra2}
 \mathcal{D}(A) B &=& A B A^{\dagger} - \frac{1}{2}\{ A^{\dagger}A, B \},
 \end{eqnarray}
where  $\{ \cdot , \cdot \}$ is the anticommutator. In (\ref{EcMaestraFinalb}) and (\ref{EcMaestra2}) $A$ and $B$ are any two linear operators. The Hamiltonian $H_{q}$ is
\begin{eqnarray}
\label{EcMaestraFinal2}
\frac{1}{\hbar} H_{q} 
&=& \delta_{1} S_{11} - \xi S_{-}^{\dagger} S_{-} + \delta (S_{-}^{2})^{\dagger} S_{-}^{2} \cr
&& + \alpha_{0}^{*}S_{-}^{2} + \alpha_{0}(S_{-}^{2})^{\dagger} .
\end{eqnarray} 
Here $\xi$ and $\delta$ are real constants, $\alpha_{0}$ is a complex number, and $\delta_{1}$ is a real parameter that can be tuned to zero or to a positive or negative value [see Sec.~\ref{Origen}]. This master equation is an effective model that has one possible physical origin presented in Sec.~\ref{Origen}.

Observe that the Liouvillian $\mathcal{L}$ includes the dissipators $\kappa_{1}\mathcal{D}(S_{-})$ and $\kappa_{2}\mathcal{D}(S_{-}^{2})$ which respectively generalize to qutrits the single-qubit and two-qubit simultaneous decay processes mentioned in the Introduction. The linear dissipator $\kappa_{1}\mathcal{D}(S_{-})$ can be thought of describing the decay of a single qutrit from an excited level to the ground level at a rate $\kappa_{1}$, while the quadratic dissipator $\kappa_{2}\mathcal{D}(S_{-}^{2})$ can be thought of as describing the simultaneous decay of two qutrits from an excited level (not necessarily the same one) to the ground level at a rate $\kappa_{2}$.  Also, the Hamiltonian $H_{q}$ includes the term $S_{11}$ which counts the number qutrits in the ground level $\vert 1 \rangle$, the quadratic term $S_{-}^{\dagger}S_{-}$ which describes the simultaneous transitions of a qutrit from an excited level to the ground level and of a qutrit from the ground level to an excited level, and the quartic term $(S_{-}^{2})^{\dagger}S_{-}^{2}$ which describes the simultaneous transitions of two qutrits from an excited level (not necessarily the same one) to the ground level and of two qutrits from the ground level to an excited level (not necessarily the same one). Finally, $H_{q}$ also includes the terms   $[ \alpha_{0}^{*}S_{-}^{2} + \alpha_{0}(S_{-}^{2})^{\dagger} ]$ which generalize to qutrits the simultaneous two-qubit driving. This coherent two-qutrit driving describes two-qutrit simultaneous transitions from the ground level to an excited level (not necessarily the same one) and viceversa.

\section{Parity symmetry of the master equation}

Consider the unitary operator associated with the parity of the number of qutrits that are found in the ground level $\vert 1 \rangle$:
\begin{eqnarray}
\label{pi}
\Pi_{0} &=& e^{i\pi S_{11}} \ = \ (-1)^{N}e^{i \pi (S_{22} + S_{33})} .
\end{eqnarray}
In the second equality in (\ref{pi}) we used $S_{11} + S_{22} + S_{33} = N$. Since $A \vert n_{1}, n_{2}, n_{3} \rangle = (-1)^{n_{1}} \vert n_{1}, n_{2}, n_{3} \rangle$ for $A = \Pi_{0}$ and $\Pi_{0}^{\dagger}$, it follows that $\Pi_{0}$ is Hermitian and that its eigenvalues are $\pm 1$. Whenever a ket $\vert \psi \rangle$ is an eigenvector of $\Pi_{0}$ we shall say that $\vert \psi \rangle$ has \textit{the parity symmetry}. Observe that $\vert \psi \rangle$ can be expressed as a linear combination of occupation number states $\vert n_{1},n_{2},n_{3} \rangle$ where $n_{1}$ is always even (odd) if $\vert \psi \rangle$ is an eigenvector of $\Pi_{0}$ with corresponding eigenvalue $+1$ (-1).

Using how $\Pi_{0}$ and the operators $S_{-}$ and $S_{-}^{\dagger}$ act on the kets of $\beta_{q}$ it is straightforward to show that
\begin{eqnarray}
\label{pi2}
[ \Pi_{0} , S_{-}^{2 }] &=& [ \Pi_{0} , (S_{-}^{\dagger})^{2} ] = [ \Pi_{0} , S_{-}^{\dagger}S_{-} ] = 0 , \cr
[ \Pi_{0} , S_{11}] &=& [ \Pi_{0} , (S_{-}^{2})^{\dagger}S_{-}^{2}] = 0, \cr
\{ \Pi_{0} , S_{-} \} &=& \{ \Pi_{0} , S_{-}^{\dagger} \} = 0 .
\end{eqnarray} 
As a consequence of (\ref{pi2}), $[ \Pi_{0}, H_{q}] = 0$.

Now define the superoperator $\mathcal{P}$ by $\mathcal{P}A = \Pi_{0} A \Pi_{0}$ for every linear operator $A$. Using that $\Pi_{0}$ is Hermitian and unitary and the properties in (\ref{pi2}) it immediately follows that $\mathcal{P}^{2} = \mathcal{I}$ with $\mathcal{I}$ the identity superoperator and that $\mathcal{P}\mathcal{L}\mathcal{P} = \mathcal{L}$. Then $\mathcal{L}\mathcal{P} = \mathcal{P}\mathcal{L}$ and, consequently,
\begin{eqnarray}
\label{pi3}
\frac{d}{dt} \rho (t) = \mathcal{L}\rho (t) \ \ &\Leftrightarrow& \ \ \frac{d}{dt} \mathcal{P}\rho (t) = \mathcal{L} [ \mathcal{P}\rho (t) ] .
\end{eqnarray}
If $\rho (0) = \mathcal{P}\rho (0)$, then the existence and uniqueness theorem for the solution of a system of linear ordinary differential equations guarantees that $\rho (t) = \mathcal{P}\rho (t)$ for all $t$. In this case $\rho(t)$ and $\Pi_{0}$ commute and the eigenvectors of $\rho (t)$ can be chosen to have the parity symmetry.

\section{Parameter independent stationary states}
\label{solEst}

We are interested in determining stationary states $\rho_{s}$ of the master equation in (\ref{EcMaestraFinal}) that are independent of the parameters. Then, $\rho_{s}$ must satisfy the conditions 
\begin{eqnarray}
\label{CondicionSS1}
\mathcal{D}(S_{-})\rho_{s} &=& 0, \quad\quad [S_{-}^{\dagger} S_{-}, \rho_{s}] = 0,  \cr
\mathcal{D}(S_{-}^{2})\rho_{s} &=& 0, \quad\quad [(S_{-}^{2})^{\dagger} S_{-}^{2}, \rho_{s}] = 0, \cr
[S_{11}, \rho_{s}] &=&  0, \quad\quad [\alpha_{0}^{*}S_{-}^{2} + \alpha_{0}(S_{-}^{2})^{\dagger}, \rho_{s}] =  0.
\end{eqnarray}
Notice that, in essence, we are \textit{looking} for \textit{dark states} of the Liouvillian $\mathcal{L}$ \cite{DarkMuller,DarkSieberer}.

Consider a pure stationary state $\rho_{s} = \vert \Phi \rangle\langle \Phi \vert$. In Appendix \ref{AppendixA} it is shown that $\rho_{s}$ satisfies (\ref{CondicionSS1}) if and only if $S_{-}\vert \Phi \rangle = (S_{-}^{2})^{\dagger} \vert \Phi \rangle = 0$ and $\vert \Phi \rangle$ is an eigenvector of $S_{11}$.  In addition, it is also proved that $S_{-}\vert \Phi \rangle = (S_{-}^{2})^{\dagger} \vert \Phi \rangle = 0$ and $\vert \Phi \rangle$ is an eigenvector of $S_{11}$ if and only if $\vert \Phi \rangle$ is equal (except for a global phase factor) to one of the two following normalized states:
\begin{eqnarray}
\label{edoEstacionario1}
\vert 0_{L} \rangle  &=& \vert z_{1}=0,z_{2}=1,z_{3}=-1\rangle_{N} , \cr
&=& \sum_{n_{3}=0}^{N} c_{n_{3}} \vert 0, N-n_{3}, n_{3} \rangle , \cr
\vert 1_{L} \rangle  &=& b_{1}^{\dagger} \vert z_{1}=0,z_{2}=1,z_{3}=-1\rangle_{N-1} , \cr
&=& \sum_{n_{3}=0}^{N-1} d_{n_{3}} \vert 1, N-1-n_{3}, n_{3} \rangle , \cr
&&
\end{eqnarray} 
where
\begin{eqnarray}
\label{edoEstacionario2}
c_{n_{3}}
&=&  (-1)^{n_{3}} \sqrt{\frac{N!}{2^{N} \ (N-n_{3})! \ n_{3}!}} \ , \cr
d_{n_{3}}
&=&  (-1)^{n_{3}} \sqrt{\frac{(N-1)!}{2^{N-1} \ (N-1-n_{3})! \ n_{3}!}} \ ,
\end{eqnarray}
and 
\begin{eqnarray}
\label{coherent1}
\vert z_{1}, z_{2}, z_{3} \rangle_{M} &=& \frac{1}{\sqrt{M!}} \left( \frac{z_{1}b_{1}^{\dagger} + z_{2}b_{2}^{\dagger} + z_{3}b_{3}^{\dagger}}{\sqrt{ \vert z_{1} \vert^{2} + \vert z_{2} \vert^{2} + \vert z_{3} \vert^{2}}} \right)^{M} \vert \mathbf{0} \rangle , \quad\quad
\end{eqnarray}
represents a coherent state of U(3) for $M$ particles \cite{Calixto21b}. Here $\vert \mathbf{0} \rangle = \vert n_{1} = 0, n_{2} = 0, n_{3} = 0 \rangle$ is the Fock vacuum and $z_i$ $(i=1,2,3)$ are complex numbers that are not all zero. In addition, note that $\vert 0_{L} \rangle$ and $\vert 1_{L} \rangle$ are orthogonal to each other because they are eigenvectors of $S_{11}$ with different eigenvalues. Note that, for convenience, we use a nonconventional, redundant representation for coherent states of U(3) and that the usual one is obtained by putting $z_{1} = 1$ \cite{Calixto21b}: according to (\ref{coherent1}) one has $\vert qz_{1}, qz_{2}, qz_{3} \rangle_{M} = (q/\vert q \vert)^{M} \vert z_{1},z_{2},z_{3} \rangle_{M}$ for any nonzero complex number $q$, so $\vert qz_{1}, qz_{2}, qz_{3} \rangle_{M}$ and $\vert z_{1},z_{2},z_{3} \rangle_{M}$ differ by a global phase factor.

We emphasize that
\begin{eqnarray}
\label{propElogicos}
S_{11} \vert 0_{L} \rangle &=& 0 , \quad\quad \ S_{-}^{\dagger}\vert 0_{L}\rangle = 0, \quad \Pi_{0} \vert 0_{L} \rangle = \vert 0_{L} \rangle , \cr
S_{11} \vert 1_{L} \rangle &=& \vert 1_{L} \rangle , \quad S_{-}^{\dagger} \vert 1_{L} \rangle \not= 0, \quad \Pi _{0}\vert 1_{L} \rangle = -\vert 1_{L} \rangle , \cr
S_{-}\vert 0_{L} \rangle &=& 0 , \quad  (S_{-}^{2})^{\dagger} \vert 0_{L} \rangle = 0 , \quad \langle \mathbb{j}_{L} \vert \mathbb{k}_{L} \rangle = \delta_{\mathbb{j}\mathbb{k}} , \cr
S_{-}\vert 1_{L} \rangle &=& 0 , \quad  (S_{-}^{2})^{\dagger} \vert 1_{L} \rangle = 0  \,,
\end{eqnarray}
where lowercase bold letters, like $\mathbb{j}$ or $\mathbb{k}$ (or $\mathbb{m}$ and $\mathbb{n}$ later), represent binary digits, i.e. $\mathbb{j},\mathbb{k}\in \{0,1\}$.
Observe that $\vert 0_{L} \rangle$ is a coherent state, while $\vert 1_{L} \rangle$ is a \textit{particle-added coherent state}, to use the same nomenclature as \cite{Agarwal} but for \textit{photon-added coherent states}. Moreover, it is straightforward to show that
\begin{eqnarray}
\label{GATOS}
\vert \mathcal{\mathbb{j}}_{L} \rangle &=& \lim_{\epsilon \rightarrow 0} \frac{1}{2N_{\mathbb{j}}(\epsilon)} \Bigg[ \vert z_{1} = \epsilon, z_{2} = 1, z_{3} = -1 \rangle_{N} \cr
&& \quad + (-1)^{\mathbb{j}} \vert z_{1} = -\epsilon , z_{2} = 1, z_{3} = -1 \rangle_{N} \Bigg] , \cr
N_{\mathbb{j}}(\epsilon) &=& \frac{1}{\sqrt{2}}\left[ 1 + (-1)^{\mathbb{j}} \left( \frac{2-\epsilon^{2}}{2+ \epsilon^{2}} \right)^{N} \right]^{1/2}  . \quad \
\end{eqnarray}
Therefore, $\vert 0_{L} \rangle$ and $\vert 1_{L} \rangle$ are Schr\"{o}dinger cat states. From (\ref{propElogicos}) notice that both $\vert 0_{L} \rangle$ and $\vert 1_{L} \rangle$ have the parity symmetry. The former is an even cat state, while the latter is an odd cat state. From the discussion above one concludes that $\vert 0_{L} \rangle$ and $\vert 1_{L} \rangle$ are highly robust even and odd Schr\"{o}dinger cat states that can be used as logical states of a qubit.

The density operators $\rho_{0} = \vert 0_{L} \rangle\langle 0_{L} \vert$ and $\rho_{1} = \vert 1_{L} \rangle\langle 1_{L} \vert$ are the only parameter independent, pure stationary states of the master equation in (\ref{EcMaestraFinal}). We emphasize that they are fixed points of the Liouvillian $\mathcal{L}$ that commute with the Hamiltonian $H_{q}$ and that satisfy $L\rho_{\scriptscriptstyle{\mathbb{j}}} = 0$ $(\mathbb{j}=0,1)$ for all the Lindblad operators $L = S_{-}$ and $S_{-}^{2}$ appearing in $\mathcal{L}$. Thus, they constitute \textit{dark states} of $\mathcal{L}$ \cite{DarkSieberer} and are immune to both the linear $\kappa_{1} \mathcal{D}(S_{-})$ and quadratic $\kappa_{2} \mathcal{D}(S_{-}^{2})$ dissipators. In addition, the properties in (\ref{propElogicos}) guarantee that they would also be immune to other interactions. For example, they are immune to those described by Hamiltonians with terms of the form $ABC$ where $A= (S_{-})^{\dagger}$ or $S_{-}^{2}$, $C = S_{-}$ or $(S_{-}^{2})^{\dagger}$, and $B$ is any linear operator. They would also be immune to dissipators $\mathcal{D}(A)$ where the linear operator $A$ is of the form $A = BC$ where $C = S_{-}$ or $(S_{-}^{2})^{\dagger}$ and $B$ is any linear operator. One could also take a convex combination of $\rho_{0}$ and $\rho_{1}$ to obtain a mixed stationary state that is independent of the parameters. 

It is important to note that there are other parameter dependent stationary solutions of the master equation in (\ref{EcMaestraFinal}). If one uses the basis $\beta_{q}$ to express $\mathcal{L} \rho = 0$ as a matrix equation and one takes $\delta_{1} = 0$, our numerical calculations indicate that the subspace of matrices that satisfy $\mathcal{L} \rho = 0$ has dimension $N+3$. In this case, an orthonormal basis for that subspace could be constructed by starting from the matrices associated with $\vert \mathbb{j}_{L} \rangle \langle \mathbb{k}_{L} \vert$ with $\mathbb{j},\mathbb{k} = 0,1$ and using the Hilbert-Schmidt inner product. Notice that $\{  \vert 0_{L} \rangle , \ \vert 1_{L} \rangle  \}$ constitutes a \textit{decoherence-free} subspace \cite{symmetry} only when $\delta_{1} = 0$. 

Finally, since  $\vert 0_{L} \rangle$ and $\vert 1_{L} \rangle$ are eigenvectors of $S_{11}$ with corresponding eigenvalues $0$ and $1$ [see (\ref{propElogicos})], they have a defined number of qutrits in the level $\vert 1 \rangle$ equal to $0$ and $1$, respectively. Therefore, the logical qubit states $\vert 0_{L} \rangle$ and $\vert 1_{L} \rangle$ cannot be obtained by using a Holstein-Primakoff expansion \cite{Klein} where the majority of the qutrits are found in the level $\vert 1 \rangle$ (the so called \textit{low-excitation regime}). This is a notable  difference with the case of qubits where the aforementioned approximation can be applied \cite{NoriO,squeezing,Nori2024}.

\subsection{Coherent state creation operators}

Since $S_{-} = S_{12} + S_{13} = b_{1}^{\dagger}(b_{2} + b_{3})$, one can take inspiration from the treatment of the isotropic two-dimensional harmonic oscillator in terms of creation and annihilation operators of left and right circular quanta \cite{Cohen1} to give a simple and elegant description of both the model and the logical states $\vert 0_{L} \rangle$ and $\vert 1_{L} \rangle$.

Define the operators
\begin{eqnarray}
\label{RL1}
b_{l} &=& \frac{1}{\sqrt{2}} (b_{2} + b_{3}) , \quad b_{r} = \frac{1}{\sqrt{2}} (b_{2} - b_{3}) .
\end{eqnarray}
Note that $[b_{l},b_{r}] = [b_{l}, b_{r}^{\dagger}] = 0$ and $[b_{l},b_{l}^{\dagger}] = [b_{r},b_{r}^{\dagger}] = 1$. Then, $b_{l}^{\dagger}$, $b_{l}$ and $b_{r}^{\dagger}$, $b_{r}$ are two pairs of creation and annihilation operators of two independent \textit{left}- and \textit{right-handed} harmonic oscillators. Also, $b_{1}$ and $b_{1}^{\dagger}$ commute with $b_{l}$, $b_{l}^{\dagger}$, $b_{r}$, and $b_{r}^{\dagger}$ and $S_{22} + S_{33} = (b_{2}^{\dagger}b_{2} + b_{3}^{\dagger}b_{3}) = (b_{l}^{\dagger}b_{l} + b_{r}^{\dagger}b_{r})$. 

Using these operators one has $S_{-} = \sqrt{2} b_{1}^{\dagger}b_{l}$, so the Liouvillian $\mathcal{L}$ in (\ref{EcMaestraFinalb})  can be expressed as
\begin{eqnarray}
\label{RL2}
\mathcal{L}A &=& -\frac{i}{\hbar} \left[ H_{q}, \ A \right] + 2\kappa_{1} \mathcal{D}(b_{1}^{\dagger}b_{l})A + 4 \kappa_{2} \mathcal{D}\left[  (b_{1}^{\dagger})^{2}b_{l}^{2} \right] A, \cr
&&
\end{eqnarray}
where $A$ is any linear operator and
\begin{eqnarray}
\label{RL3}
\frac{1}{\hbar} H_{q} &=& \delta_{1}S_{11} - 2 \xi  (S_{11} + 1) b_{l}^{\dagger}b_{l} + 4 \delta b_{1}^{2} (b_{1}^{\dagger})^{2} (b_{l}^{\dagger})^{2} b_{l}^{2} \cr
&& \quad + 2\alpha_{0}^{*}(b_{1}^{\dagger})^{2}b_{l}^{2} + 2 \alpha_{0} b_{1}^{2}(b_{l}^{\dagger})^{2} .
\end{eqnarray}
Observe that $b_{r}^{\dagger}$ and $b_{r}$ do not appear in the Liouvillian $\mathcal{L}$. This is the key to finding the parameter independent, pure stationary states by mere inspection. In fact, 
$b_{r}^{\dagger}$ and $b_{r}$ are associated with the \textit{dark degrees of freedom}, while $b_{l}^{\dagger}$ and $b_{l}$ are associated with \textit{bright degrees of freedom}.

Another orthonormal basis for the state space of the $N$ qutrits can be obtained by using the occupation number states defined by $b_{1}^{\dagger}$, $b_{l}^{\dagger}$, and $b_{r}^{\dagger}$: 
\begin{eqnarray}
\label{RL4}
\beta_{q} ' &=& \Bigg\{ \ \vert n_{1}, n_{l}, n_{r} \rangle = \frac{(b_{1}^{\dagger})^{n_{1}}}{\sqrt{n_{1}!}} \frac{(b_{l}^{\dagger})^{n_{l}}}{\sqrt{n_{l}!}} \frac{(b_{r}^{\dagger})^{n_{r}}}{\sqrt{n_{r}!}}  \vert  \mathbf{0} \rangle : \cr
&& \quad\quad n_{1},n_{l},n_{r} \in \mathbb{Z}^{+}, \ n_{1} + n_{l} + n_{r} = N \ \Bigg\} . \quad\quad
\end{eqnarray}
Then, for $k=1,l,r$ one has
\begin{eqnarray}
\label{RL5}
b_{k}^{\dagger}\vert n_{1}, n_{l}, n_{r} \rangle &=&  \sqrt{n_{k}+1}\vert n_{1}+\delta_{k,1}, n_{l}+\delta_{k,l}, n_{r} + \delta_{k,r} \rangle , \cr
b_{k}\vert n_{1}, n_{l}, n_{r} \rangle &=&  \sqrt{n_{k}}\vert n_{1}-\delta_{k,1}, n_{l} -\delta_{k,l}, n_{r} -\delta_{k,r} \rangle . \quad\quad
\end{eqnarray}
Note that, using the definition of $b_{l}$ and $b_{r}$ in (\ref{RL1}) and the definition of coherent states in (\ref{coherent1}), one has 
\begin{eqnarray}
\label{RL6}
\vert n_{1}, n_{l}, n_{r} = 0 \rangle &=& \frac{(b_{1}^{\dagger})^{n_{1}}}{\sqrt{n_{1}!}} \frac{(b_{l}^{\dagger})^{n_{l}}}{\sqrt{n_{l}!}} \vert \mathbf{0} \rangle  , \cr
&= & \frac{(b_{1}^{\dagger})^{n_{1}}}{\sqrt{n_{1}!}}  \vert z_{1} = 0, z_{2} = 1 , z_{3} = 1\rangle_{n_{l}} , \cr
\vert n_{1}, n_{l} = 0, n_{r} \rangle &=& \frac{(b_{1}^{\dagger})^{n_{1}}}{\sqrt{n_{1}!}} \frac{(b_{r}^{\dagger})^{n_{r}}}{\sqrt{n_{r}!}} \vert \mathbf{0} \rangle , \cr
&= & \frac{(b_{1}^{\dagger})^{n_{1}}}{\sqrt{n_{1}!}}  \vert z_{1} = 0, z_{2} = 1 , z_{3} = -1\rangle_{n_{r}} , \quad\quad
\end{eqnarray}
with $n_{1} + n_{l} + n_{r}= N$ and $n_{1},n_{l},n_{r} \in \mathbb{Z}^{+}$. Therefore, $\vert n_{1}, n_{l}, n_{r} = 0 \rangle$ and $\vert n_{1}, n_{l} = 0, n_{r} \rangle$ are coherent states of $n_{l}$ and $n_{r}$ particles to which one adds $n_{1}$ particles in the level $\vert 1  \rangle$, respectively. Taking $n_{1} = 0$ one finds that $b_{l}^{\dagger}$ and $b_{r}^{\dagger}$ are creation operators of coherent states of U(3).

From (\ref{edoEstacionario1}) and (\ref{RL6}) one finds that the logical qubit states can be expressed simply as
\begin{eqnarray}
\label{RL7}
\vert 0_{L} \rangle &=& \vert n_{1} = 0, n_{l} = 0, n_{r} = N \rangle  , \cr
\vert 1_{L} \rangle &=& \vert n_{1} = 1, n_{l} = 0, n_{r} = N-1 \rangle . 
\end{eqnarray}
From (\ref{RL3}) and (\ref{RL4}) one could have easily deduced by mere inspection that the logical states in (\ref{RL7}) satisfy $\mathcal{L}\vert \mathbb{k}_{L} \rangle\langle \mathbb{k}_{L} \vert = 0$ with $\mathbb{k}=0,1$. First observe that the expression for $\mathcal{L}$ does not include the operators $b_{r}$ and $b_{r}^{\dagger}$, so one would propose as a stationary solution a pure state of the form $\vert n_{1}, n_{l}, n_{r} \rangle\langle n_{1}, n_{l}, n_{r} \vert$ with $n_{1} + n_{l} + n_{r} = N$.  Choosing $n_{l} = 0$ will eliminate the dissipators and all the terms of the commutator involving $H_{q}/\hbar$ except those that are multiplied by $\delta_{11}$, $\alpha_{0}$, and $\alpha_{0}^{*}$. Since the terms that are multiplied by $\alpha_{0}$ and $\alpha_{0}^{*}$ involve either $(b_{1}^{\dagger})^{2}b_{l}^{2}$ or $b_{1}^{2}(b_{l}^{\dagger})^{2}$, they disappear by choosing $n_{1} = \mathbb{k}$ and $n_{l} = 0$ with $\mathbb{k} = 0$ or $1$ but this choice of $n_{1}$ also eliminates the term multiplied $\delta_{1}$. Therefore, one arrives to the pure stationary states defined by (\ref{RL7}).

\subsection{Collective dephasing}

In this section we first discuss the resilience of the logical qubit states $\vert 0_{L} \rangle$ and $\vert 1_{L} \rangle$ to pure collective dephasing processes, that is, the damping of coherences without changing the populations of the levels of the qutrits. The description of pure dephasing in qutrits is not as straightforward as it is for qubits \cite{EITreview,Dephasing1,DephasingJCS,Dephasing2,Dephasing3}. We consider two types of pure collective dephasing described by 
\begin{eqnarray}
\label{Dephasing1}
\mathcal{L}_{\mathrm{cd}}A &=&  \Gamma_{1} \mathcal{D}(S_{11})A + \Gamma_{23} \mathcal{D}(S_{22} + S_{33})A, \cr
\mathcal{L}_{\mathrm{ud}}A &=&  \sum_{j=1,2,3} \Gamma_{j} \mathcal{D}(S_{jj})A ,
\end{eqnarray}
where $A$ is any linear operator. Here $\mathcal{L}_{\mathrm{ud}}$ describes uncorrelated collective dephasing, while the term $\mathcal{D}(S_{22} + S_{33})$ in $\mathcal{L}_{\mathrm{cd}}$ models correlated collective dephasing in the levels $\vert 2 \rangle$ and $\vert 3 \rangle$.

Since the logical qubit states $\vert 0_{L} \rangle$ and $\vert 1_{L} \rangle$ are eigenvectors of $S_{11}$ and $(S_{22} + S_{33}) = (N - S_{11})$ [see (\ref{propElogicos})], it is straightforward to show that $\mathcal{L}_{\mathrm{cd}}\vert \mathbb{j}_{L} \rangle\langle \mathbb{j}_{L} \vert = 0$ for $\mathbb{j}=0,1$. Therefore, the logical qubit states are immune to the pure collective dephasing described by $\mathcal{L}_{\mathrm{cd}}$. 

On the other hand, the logical qubit states $\vert 0_{L} \rangle$ and $\vert 1_{L} \rangle$ are vulnerable to the pure uncorrelated, collective dephasing described by $\mathcal{L}_{\mathrm{ud}}$ because $\mathcal{L}_{\mathrm{ud}}\vert \mathbb{j}_{L} \rangle\langle \mathbb{j}_{L} \vert \not= 0$. Its effects along with those of local dephasing are presented for two qutrits in Sec.~\ref{2qutrits}.
  
\section{Quantum gates}

In this section we show how to implement a quantum NOT gate and a Hadamard gate $\mathcal{H}$ followed by either the phase gate $\mathcal{S}$ or the phase $\mathcal{S}$ and $\mathcal{Z}$ gates \cite{Chuang}. All of these can be carried out simply by tuning $\delta_{1}$ to zero and nonzero real values. This tuning corresponds to adjusting the frequency of a laser to be resonant or not resonant with twice the qutrits transition frequency $\omega_{q}$ [see Sec.~\ref{Origen}]. We assume that one can prepare the ensemble of qutrits in one of the states $\vert \mathbb{j}_{L}, \varphi \rangle$ defined below.

For any $\varphi \in (-\pi, \pi]$ consider the states
\begin{eqnarray}
\label{L0}
\vert \mathbb{j}_{L}, \varphi \rangle &=& \frac{1}{\sqrt{2}}\left[ \ \vert 0_{L} \rangle + (-1)^{\mathbb{j}} e^{i\varphi} \vert 1_{L} \rangle \ \right]  . \quad\quad
\end{eqnarray}
Observe that $\vert 0_{L}, \varphi \rangle$ and $\vert 1_{L}, \varphi \rangle$ are normalized and orthogonal to each other. From (\ref{propElogicos}) it is clear that $\vert \mathbb{j}_{L}, \varphi \rangle \langle \mathbb{j}_{L}, \varphi \vert$  are stationary density operators of the master equation (\ref{EcMaestraFinal}) if and only if $\delta_{1} = 0$. Hence, $\vert 0_{L}, \varphi \rangle$ and $\vert 1_{L}, \varphi \rangle$  can also be used as robust, parameter independent logical states of a qubit when $\delta_{1} = 0$. In the next section we discuss how to prepare these states in the case of two qutrits.

First assume that $\delta_{1} = 0$ and that the state of the system is $\rho_{q}(0) = \vert \mathbb{j}_{L}, \varphi \rangle\langle \mathbb{j}_{L},\varphi \vert$ with $\mathbb{j}=0$ or $1$. Then, the system is found in a stationary state and we want to apply either a quantum NOT gate, a $\mathcal{SH}$ transformation, or a $\mathcal{ZSH}$ operation. In order to do this we now tune $\delta_{1}$ to any nonzero value. As a consequence, $\rho_{q}(0)$ is no longer a stationary state and it will evolve. 

Take
\begin{eqnarray}
\label{L1}
\rho_{q} (t) 
&=& \sum_{\mathbb{m},\mathbb{n}} A_{\mathbb{m}\mathbb{n}}(t) \vert \mathbb{m}_{L}, \varphi \rangle\langle \mathbb{n}_{L}, \varphi \vert ,
\end{eqnarray}
where $A_{\mathbb{m}\mathbb{n}}(t)$ are complex-valued functions. Using (\ref{propElogicos}) it is straightforward to show that
\begin{eqnarray}
\label{L2}
\mathcal{L} \rho_{q}(t)
&=& -i \frac{\delta_{1}}{2} (-1)^{N}
\Bigg\{ 
\left[ A_{01}(t) - A_{10}(t) \right] \vert 0_{L},\varphi \rangle\langle 0_{L}, \varphi \vert \cr
&& \quad + \left[ A_{00}(t) - A_{11}(t) \right] \vert 0_{L}, \varphi \rangle\langle 1_{L}, \varphi \vert \cr 
&& \quad + \left[ A_{11}(t) - A_{00}(t) \right] \vert 1_{L}, \varphi \rangle\langle 0_{L}, \varphi \vert \cr
&& \quad + \left[ A_{10}(t) -A_{01}(t) \right] \vert 1_{L}, \varphi \rangle\langle 1_{L}, \varphi \vert  \Bigg\} .
\end{eqnarray}
Substituting (\ref{L1}) and (\ref{L2}) into the master equation (\ref{EcMaestraFinal}) one obtains a linear system of ordinary differential equations that can be solved exactly:
\begin{eqnarray}
\label{L3}
\mathbf{A}(t) &=&  
\left(
\begin{array}{cc}
\mathbb{B}_{0}( \delta_{1}t /2 ) & \mathbb{B}_{1}(t) \cr
\mathbb{B}_{1}(t) & \mathbb{B}_{0}( \delta_{1}t /2 )
\end{array}
\right) \mathbf{A}(0),
\end{eqnarray}
where
\begin{eqnarray}
\label{L4}
\mathbf{A}(t) &=& \left( A_{00}(t), A_{11}(t), A_{01}(t), A_{10}(t) \right)^{T} \ , \cr
\mathbb{B}_{0}(\tau) &=& 
\left(
\begin{array}{cc}
\mathrm{cos}^{2}(\tau) & \mathrm{sin}^{2}(\tau) \cr
\mathrm{sin}^{2}(\tau) & \mathrm{cos}^{2}(\tau)
\end{array}
\right) , \cr
\mathbb{B}_{1}(t) &=& 
- \frac{i}{2} \mathrm{sin}(\delta_{1}t) \left(
\begin{array}{cc}
1 & -1 \cr
-1 & 1
\end{array}
\right) .
\end{eqnarray}
The superscript $T$ denotes the transpose. 

Assume $A_{\mathbb{m}\mathbb{n}}(0) = \delta_{\mathbb{m}\mathbb{j}}\delta_{\mathbb{n}\mathbb{j}}$. Then one can calculate $\rho_{q}(t)$ using (\ref{L1})-(\ref{L4}). In particular, $\rho_{q}(0) = \vert \mathbb{j}_{L},\varphi \rangle\langle \mathbb{j}_{L},\varphi \vert$ and  one can produce oscillations between $\vert 0_{L}, \varphi \rangle$ and $\vert 1_{L}, \varphi \rangle$ by letting $\delta_{1} \not= 0$.

If one tunes $\delta_{1}$ to zero at $\delta_{1}t_{p} = \pi(2p+1)$ with $p$ an integer, then
\begin{eqnarray}
\label{L5}
\rho_{q}(t_{p}) &=& \delta_{1\mathbb{j}} \vert 0_{L},\varphi \rangle\langle 0_{L}, \varphi \vert + \delta_{0\mathbb{j}} \vert 1_{L}, \varphi \rangle\langle 1_{L}, \varphi \vert  , \cr
&=& \vert (\mathbb{j} \oplus 1)_{L},\varphi \rangle\langle (\mathbb{j} \oplus 1)_{L},\varphi \vert .
\end{eqnarray} 
Here $\oplus$ denotes modulo $2$ addition. Hence, one performs the transformation 
\[
\vert \mathbb{j}_{L}, \varphi \rangle\langle \mathbb{j}_{L},\varphi \vert \rightarrow \vert (\mathbb{j} \oplus 1)_{L},\varphi \rangle\langle (\mathbb{j} \oplus 1)_{L}, \varphi \vert ,
\] 
and one has implemented a quantum NOT gate.

If instead one tunes $\delta_{1}$ to zero at $\delta_{1}t_{p} = \pi(2p+1)/2$ with $p$ an integer, then 
\begin{eqnarray}
\label{L6}
\rho_{q}(t_{p}) &=& 
\left\{
\begin{array}{cc}
\vert \Psi_{+p} \rangle\langle \Psi_{+p} \vert  & \mbox{if $\mathbb{j}=0$}, \cr
\vert \Psi_{-p} \rangle\langle \Psi_{-p} \vert  & \mbox{if $\mathbb{j}=1$}.
\end{array}
\right.
\end{eqnarray}
with
\begin{eqnarray}
\label{L7}
\vert \Psi_{\pm p} \rangle &=& \frac{1}{\sqrt{2}} \Bigg[ \vert 0_{L}, \varphi \rangle \pm i (-1)^{p}\vert 1_{L}, \varphi \rangle \Bigg] . 
\end{eqnarray}
Hence, one has implemented a $\mathcal{SH}$ gate if $p$ is even or a $\mathcal{ZSH}$ gate if $p$ is odd.

\section{The case of two qutrits}
\label{2qutrits}

In this and only this section we assume that $N=2$. We first show how the logical qubit states can be prepared, we construct an orthogonal basis for the vector space of stationary solutions of the master equation in (\ref{EcMaestraFinal}), and then we present the effects of uncorrelated pure collective dephasing as described by $\mathcal{L}_{\mathrm{ud}}$ in (\ref{Dephasing1}), as well as those of inhomogeneous broadening and local dephasing.

Assume that the kets of $\beta_{q}$ are ordered as follows:
\begin{eqnarray}
\label{N1}
\beta_{q} &=& \Big\{ \ \vert 2,0,0 \rangle, \ \vert 1,1,0 \rangle, \ \vert 0,2,0 \rangle , \ \vert 1,0,1 \rangle , \cr
&& \quad\quad \vert 0,1,1 \rangle , \ \vert 0,0,2 \rangle \ \Big\} .
\end{eqnarray}
Here the ordered triple of numbers inside the ket symbol corresponds to the ordered triple $(n_{1},n_{2},n_{3})$.
Notice that the state space of the qutrits has dimension $6$ and that the logical states of the qubit take the form
\begin{eqnarray}
\label{N2}
\vert 0_{L} \rangle &=& \frac{1}{2} \left( \vert 0,2,0 \rangle -\sqrt{2}\vert 0,1,1 \rangle + \vert 0,0,2 \rangle \right) , \cr
\vert 1_{L} \rangle &=& \frac{1}{\sqrt{2}} \left( \ \vert 1,1,0 \rangle - \vert 1,0,1 \rangle \ \right) .
\end{eqnarray}
Also, the matrix representation of a linear operator $A$ with respect to the basis $\beta_{q}$ is a $6\times 6$ complex matrix denoted by $[A]_{\beta}$. 

\subsection{A possible preparation of the logical states}

One can prepare the system of two qutrits in the logical states $\vert 0_{L} \rangle$ and $\vert 1_{L} \rangle$ by means of a coherent external driving. In the context of the physical system of Sec.~\ref{Origen}, this would be performed before coupling the qutrits to the rest of the subsystems. 

If the two qutrits evolve as a closed system under the Hamiltonian
\begin{eqnarray}
\label{N9}
H_{D1}(t) &=& \hbar g_{d} (S_{13} + S_{31}) - \hbar g_{d} (S_{12} + S_{21}) ,
\end{eqnarray}
with $g_{d} >0$ and the initial state is the Fock state $\vert 2,0,0 \rangle$ (that is, the two qutrits are initially in their respective ground levels), then with probability $1$ one finds the system of two qutrits in the state $\vert 0_{L} \rangle$ at times $(2n+1)\pi /(2\sqrt{2} g_{d})$ with $n$ an integer. 

If instead the qutrits evolve as a closed system under the effective Hamiltonian 
\begin{eqnarray}
\label{N10}
H_{D2}(t) &=& i \hbar g_{d}(S_{23} - S_{32}) , \quad
\end{eqnarray}
with $g_{d}>0$ and the initial state is the Fock state $\vert 1,1,0 \rangle$ or the Fock state $\vert 1,0,1\rangle$ (that is, one qutrit is in the ground level while the other is in an excited level), then with probability $1$ one finds the system of two qutrits in the state $\vert 1_{L} \rangle$ at the respective times $(4n+1)\pi /(4 g_{d})$ and $(4n-1)\pi /(4 g_{d})$ with $n$ an integer.

\subsection{A basis for the subspace of stationary states}

In this and only this section we assume that $\delta_{1} = 0$ and that $\alpha_{0}$ is pure imaginary. This case is particularly attractive because all stationary solutions of the master equation (\ref{EcMaestraFinal}) have a relatively simple form.

Using the basis $\beta_{q}$ one can express $\mathcal{L}\rho = 0$ as a matrix equation. The set of all $6 \times 6$ complex matrices that satisfy this equation is a vector space $\mathcal{S}$ over the complex numbers that (according to our numerical calculations) has dimension $5$. One can equip $\mathcal{S}$ with the Hilbert-Schmidt inner product: $(\mathbb{A}, \mathbb{B}) = \mathrm{Tr}(\mathbb{A}^{\dagger} \mathbb{B})$ for any two matrices $\mathbb{A}, \mathbb{B} \in \mathcal{S}$. Whenever we say that a matrix is normalized or that two matrices are orthogonal it is to be understood that it is with respect to the Hilbert-Schmidt inner product and the associated norm. In particular, a pure state density matrix has norm $1$, while a mixed state density matrix has norm $<1$.

Since $\delta_{1} = 0$, we already know from Sec.~\ref{solEst} that $\{ \ [ \ \vert \mathbb{j}_{L} \rangle\langle \mathbb{k}_{L} \vert \ ]_{\beta_{q}} : \ \mathbb{j},\mathbb{k}=0,1 \ \}$ is an orthonormal set contained in $\mathcal{S}$. It can be extended to an orthogonal basis for $\mathcal{S}$ given by
\begin{eqnarray}
\label{N3}
\beta_{\mathcal{S}} &=& \Big\{ \ [ \ \vert \mathbb{j}_{L} \rangle\langle \mathbb{k}_{L} \vert \ ]_{\beta_{q}} : \ \mathbb{j},\mathbb{k} = 0,1 \ \Big\} \cup \Big\{  [ \rho_{00} ]_{\beta_{q}} \Big\} , \nonumber
\end{eqnarray}
where the density operator $\rho_{00}$ is defined by 
\begin{eqnarray}
\label{N4}
[ \rho_{00} ]_{\beta_{q}} &=& \rho_{R} + i\rho_{I} ,
\end{eqnarray}
where
\begin{eqnarray}
\label{N5}
\rho_{R} &=&
2 p_{3}
\left(
\begin{array}{cccccc}
0 & 0 & 0& 0 & 0 & 0 \cr
0 & 1 & 0& -1 & 0 & 0 \cr
0 & 0 & 0& 0 & 0 & 0 \cr
0 & -1 & 0& 1 & 0 & 0 \cr
0 & 0 & 0& 0 & 0 & 0 \cr
0 & 0 & 0& 0 & 0 & 0 \cr
\end{array}
\right)
\cr
&& + p_{3}
\left(
\begin{array}{cccccc}
0 & 0 & 0& 0 & 0 & 0 \cr
0 & 0 & 0& 0 & 0 & 0 \cr
0 & 0 & 1& 0 & -\sqrt{2} & 1 \cr
0 & 0 & 0& 0 & 0 & 0 \cr
0 & 0 & -\sqrt{2} & 0 & 2 & -\sqrt{2} \cr
0 & 0 & 1& 0 & -\sqrt{2} & 1 \cr
\end{array}
\right) \cr
&& + p_{2} G
\left(
\begin{array}{cccccc}
p_{1} & 0 & -F& 0 & -\sqrt{2} F & -F \cr
0    & 0 & 0& -\sqrt{2} & 0 & 0 \cr
-F  & 0 & 0& 0 & -1 & 0 \cr
0    & -\sqrt{2} & 0& 0 & 0 & 0 \cr
-\sqrt{2} F & 0 & -1 & 0 & 0 & -1 \cr
-F  & 0 & 0& 0 & -1 & 0 
\end{array}
\right)
 , \cr
\rho_{I}
&=&
p_{2}\left(
\begin{array}{cccccc}
0 & 0 & 1& 0 & \sqrt{2} & 1 \cr
0 & 0 & 0& 0 & 0 & 0 \cr
-1 & 0 & 0& 0& 0 & 0 \cr
0 & 0 & 0& 0 & 0 & 0 \cr
-\sqrt{2} & 0 & 0& 0 & 0& 0 \cr
-1 & 0 & 0& 0 & 0& 0 \cr
\end{array}
\right) , 
\end{eqnarray}
and
\begin{eqnarray}
\label{N6}
F &=& \frac{4\kappa_{2} + \kappa_{1}}{2\sqrt{2} \,\mathrm{Im}( \alpha_{0} )} , \quad\quad\quad\quad\quad\quad\quad G = \frac{\sqrt{2}\, \mathrm{Im}(\alpha_{0})}{\xi - 4\delta} , \cr
p_{1} &=& -\sqrt{2} \left( 1 +2F^{2} + \frac{2}{G^{2}} \right) , \quad p_{2} = \frac{2(4\delta - \xi )}{\mathrm{Im}(\alpha_{0})}, \cr
p_{3} &=& \frac{\mathrm{Im}(\alpha_{0})^{2}}{(4\kappa_{2} + \kappa_{1})^{2} + 4(\xi - 4\delta)^{2} + 12 \,\mathrm{Im}(\alpha_{0})^{2}} .
\end{eqnarray}
Here $\mathrm{Im}(\alpha_{0})$ denotes the imaginary part of $\alpha_{0}$.

The set $\beta_{\mathcal{S}}$ contains three density matrices, namely, those associated with the pure states $\vert 0_{L} \rangle\langle 0_{L} \vert$ and $\vert 1_{L} \rangle\langle 1_{L} \vert$ and with the mixed state $\rho_{00}$. Notice that $\rho_{00}$ is not normalized because it is a mixed state:
\begin{eqnarray}
\label{N7}
\mathrm{Tr}(\rho_{00}^{2}) &=& 1 - \frac{ 8\, \mathrm{Im}(\alpha_{0})^{2}}{(4\kappa_{2} + \kappa_{1})^{2} + 4(\xi - 4\delta)^{2} + 12\,\mathrm{Im}(\alpha_{0})^{2}} , \cr
 &<& 1. \quad
\end{eqnarray} 
Observe that $\rho_{00}$ tends to a pure state in a strongly dissipative system with $(4\kappa_{2} + \kappa_{1}) \gg 2\sqrt{2}\vert \alpha_{0} \vert$ or if $\alpha_{0} \rightarrow 0$ [the latter corresponds to no driving in the physical system of Sec.~VII].

Any stationary state of the master equation in (\ref{EcMaestraFinal}) can be expressed as a linear combination of the operators associated with the matrices in $\beta_{\mathcal{S}}$. In our numerical calculations we found that $\vert 2,0,0\rangle$ relaxes to $\rho_{00}$, that is, $e^{\mathcal{L}t} \vert 2,0,0\rangle\langle 2,0,0\vert \rightarrow \rho_{00}$ as $t\rightarrow + \infty$. Actually, one has 
\begin{eqnarray}
\label{N8}
\left(  [ \vert 2,0,0 \rangle\langle 2,0,0 \vert ]_{\beta_{q}}, [\rho_{00}]_{\beta_{q}} \right) &=& \mathrm{Tr}(\rho_{00}^{2}) ,
\end{eqnarray}
so the Cauchy-Schwarz inequality indicates that $\rho_{00} \rightarrow \vert 2,0,0 \rangle\langle 2,0,0 \vert$ if $\mathrm{Tr}(\rho_{00}^{2}) \rightarrow 1$.

\subsection{The effect of uncorrelated collective dephasing}

\begin{figure}[htbp]
    \centering
    \subfloat[]{\label{Figure3a}}\includegraphics[scale=0.9]{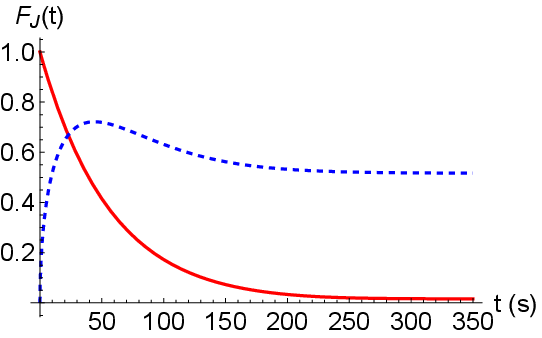}
    \subfloat[]{\label{Figure3b}}\includegraphics[scale=0.9]{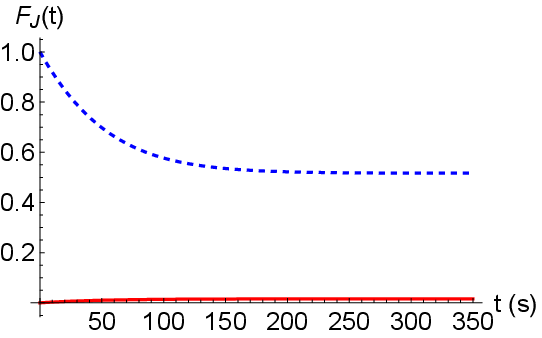}
\caption{\label{Figure3} The figures show the fidelity $F_{\mathbb{j}}(t)$ of $\rho(t)$, evolving under $\mathcal{L}+\mathcal{L}_{\mathrm{ud}}$, with $\vert 0_{L} \rangle\langle 0_{L} \vert$ (red line) and $\vert 1_{L} \rangle\langle 1_{L} \vert$ (blue-dashed line) as a function of time $t$ when $\rho(0) = \vert 0_{L} \rangle\langle 0_{L} \vert$, Fig.~\ref{Figure3a}, and when $\rho(0) = \vert 1_{L} \rangle\langle 1_{L} \vert$, Fig.~\ref{Figure3b}. In both figures the steady state values of the fidelity are $F_{0}^{\infty} = 0.016$ and $F_{1}^{\infty} = 0.52$.  The values of the parameters in both figures are $\delta_{1} = 0$, $\xi = 0.81$, $\alpha_{0} = 6.29 + i0.37$, $\delta = -3.15$, $\kappa_{1} = 7.99$, $\kappa_{2} = 0.65$, $\Gamma_{1} = 0.024$, $\Gamma_{2} = 0.032$, $\Gamma_{3} = 0.038$.}
\end{figure}

In general, uncorrelated pure collective dephasing has a detrimental effect on the logical states $\vert 0_{L} \rangle$ and $\vert 1_{L} \rangle$. Numerically we found that, if one adds $\mathcal{L}_{\mathrm{ud}}\rho(t)$ to the righthand side of the master equation (\ref{EcMaestraFinal}), then all density operators tend to a unique stationary state which has a nonnegligible fidelity with both $\vert 0_{L} \rangle\langle 0_{L} \vert$ and $\vert 1_{L} \rangle\langle 1_{L} \vert$ but the fidelity is always less than one. Fig.~\ref{Figure3} illustrates the evolution of the fidelity $F_{\mathbb{j}}(t) = \sqrt{\langle \mathbb{j}_{L} \vert \rho(t) \vert \mathbb{j}_{L} \rangle}$  of $\rho (t)$ with $\vert 0_{L} \rangle\langle 0_{L} \vert$ (red line) and $\vert 1_{L} \rangle\langle 1_{L} \vert$ (blue-dashed line) when $\rho(0) = \vert 0_{L} \rangle\langle 0_{L} \vert$, Fig.~\ref{Figure3a}, and when $\rho(0) = \vert 1_{L} \rangle\langle 1_{L} \vert$, Fig.~\ref{Figure3b}. In general, the lifetime of the logical states is $\sim \frac{4}{\Gamma_{2} + \Gamma_{3}}$.

\subsection{The effect of local dephasing and inhomogeneous broadening}

We now consider local dephasing and inhomogeneous broadening. For example, the latter can be due to Doppler shifts or the spatial inhomogeneity of the electromagnetic field which interacts with the ensemble of qutrits [see Sec.~VII], while the former can result from collisions within the ensemble. These physical processess distinguish the qutrits and, thus, break the permutation symmetry of the system and inhibit the collective behavior of the ensemble.  In order to be able to describe these processes one now has to work in the tensor product space of the two qutrits which is isomorphic to $\mathbb{C}^{3} \otimes \mathbb{C}^{3}$, has dimension $9$, and has the orthonormal basis $\beta_{tp} = \{ \vert j,k\rangle: \ j,k = 1,2,3 \}$. Here $\vert j, k \rangle = \vert j \rangle \otimes \vert k \rangle$ is the tensor product of the kets $\vert j \rangle$ and $\vert k\rangle$ where the first qutrit is in the level $\vert j\rangle$ and the second qutrit is in the level $\vert k\rangle$. In addition, all operators and the master equation in (\ref{EcMaestraFinal}) must be expressed in terms of the tensor product basis $\beta_{tp}$. In particular, we introduce the one-qutrit operators
\begin{eqnarray}
\label{one1}
\sigma_{jk} &=& \vert j \rangle\langle k \vert \quad\quad (j,k=1,2,3),
\end{eqnarray}
and their extension to the tensor product space of two qutrits
\begin{eqnarray}
\label{one2}
\sigma_{jk}^{(1)} &=& \sigma_{jk} \otimes \mathbb{I}, \quad\quad \sigma_{jk}^{(2)} = \mathbb{I} \otimes \sigma_{jk} .
\end{eqnarray}
Here $\mathbb{I}$ is the identity operator in the state space of one qutrit.
 
We have already shown that the logical qubit states are immune to global correlated dephasing as described by $\mathcal{L}_{\mathrm{cd}}$ in (\ref{Dephasing1}) but that they are vulnerable to uncorrelated collective dephasing described by $\mathcal{L}_{\mathrm{ud}}$ in (\ref{Dephasing1}). It turns out that they are also vulnerable to local uncorrelated dephasing. Therefore, we now consider correlated local dephasing and inhomogeneous broadening as described by the Liouvillian
\begin{eqnarray}
\label{IHB1}
\mathcal{L}_{\mathrm{loc}}A &=& -\frac{i}{\hbar} \left[ H_{\mathrm{B}} , \ A  \right] + \Gamma_{\mathrm{D}} \sum_{n=1}^{2} \mathcal{D} \left[ \sigma_{22}^{(n)} + \sigma_{33}^{(n)} \right] A , \quad\quad
\end{eqnarray}
where $A$ is any operator, $\Gamma_{D}>0$ is the dephasing rate, and $H_{\mathrm{B}}$ is the Hamiltonian describing inhomogeneous broadening
\begin{eqnarray}
\label{IHB3}
H_{\mathrm{B}} 
&=& \sum_{n=1}^{2} \hbar \delta\omega_{n} \left[ \sigma_{22}^{(n)}+\sigma_{33}^{(n)} \right] .
\end{eqnarray}
with $\delta\omega_{n}$ a real number. Notice that $H_{\mathrm{B}}$ describes a type of correlated inhomogeneous broadening such as that arising from Doppler shifts because the excited levels are subject to the same shift. Also, $\Gamma_{\mathrm{D}} \mathcal{D} [ \sigma_{22}^{(n)} + \sigma_{33}^{(n)} ]$ represents a correlated local dephasing where the excited levels of each qutrit are affected in the same way and both qutrits are subject to the same dephasing rate. 

It happens that the logical states $\vert 0_{L}\rangle$ and $\vert 1_{L}\rangle$ are parity-sensitive to both inhomogeneous broadening and to local correlated dephasing. It is straightforward to show that $\mathcal{L}_{\mathrm{loc}}(\vert 0_{L}\rangle\langle 0_{L} \vert) = 0$ independent of the parameters, so the logical qubit state $\vert 0_{L} \rangle$ is immune to both inhomogeneous broadening and correlated local dephasing described by $\mathcal{L}_{\mathrm{loc}}$. On the other hand, the logical qubit state $\vert 1_{L} \rangle$ is vulnerable to $\mathcal{L}_{\mathrm{loc}}$, since  $\mathcal{L}_{\mathrm{loc}}(\vert 1_{L}\rangle\langle 1_{L} \vert) \not= 0$. Nevertheless, for $\vert 1_{L} \rangle$ it is found that the inhomogeneous broadening term produces an \textit{oscillatory flux} between the symmetrical (boson) subspace and the antisymmetrical (fermion) subspace. Recall that the $9$-dimensional tensor product space of two qutrits can be decomposed as the direct
sum of the $6$-dimensional symmetric plus the $3$-dimensional antisymmetric subspaces. For $N>2$ one could talk about the \textit{superradiant} and \textit{subradiant} subspaces, like in \cite{NoriO}. The former coincides with the symmetrical subspace, while the latter includes the rest. It is important to note that the discrepancy in the resilience to inhomogeneous broadening between the even and odd cat states $\vert 0_{L} \rangle$ and $\vert 1_{L}\rangle$ encoding the qubit is much more marked in the case of an ensemble of qutrits than in the case of an ensemble of qubits, since \cite{IHBnori} found in the case of the ensemble of qubits that such discrepancy is significant only for small amplitude cat states.

One can take advantage of the \textit{oscillatory flux} to construct a superposition of symmetric and antisymmetric states that is immune to inhomogeneous broadening and, as a bonus, that is also invulnerable to correlated local dephasing. Define the antisymmetrical state
\begin{eqnarray}
\label{IHB4}
\vert 1_{L}^{A} \rangle &=& \frac{1}{\sqrt{2}} \left[ \vert 1\rangle \otimes \left( \frac{\vert 2\rangle - \vert 3\rangle}{\sqrt{2}} \right)  - \left( \frac{\vert 2\rangle - \vert 3\rangle}{\sqrt{2}} \right) \otimes \vert 1\rangle  \right] , \cr
&&
\end{eqnarray}
and the mixed permutation symmetry states
\begin{eqnarray}
\label{IHB5}
\vert \tilde{1}_{L} \rangle_{\scriptscriptstyle{\pm}} &=& \frac{1}{\sqrt{2}} \left( \vert 1_{L}\rangle \pm \vert 1_{L}^{A}\rangle  \right) = \left\{
\begin{array}{cc}
 \vert 1 \rangle \otimes \left( \frac{\vert 2\rangle - \vert 3\rangle}{\sqrt{2}} \right) & \mbox{if  $+$},\cr
\left(  \frac{\vert 2 \rangle - \vert 3 \rangle}{\sqrt{2}} \right) \otimes \vert 1 \rangle & \mbox{if  $-$ }.
\end{array}
\right. \cr
&&
\end{eqnarray}
One can show that $\mathcal{L}_{\mathrm{loc}}(\vert \tilde{1}_{L} \rangle_{\scriptscriptstyle{\pm}\scriptscriptstyle{\pm}}\langle \tilde{1}_{L} \vert ) = \mathcal{L}(\vert \tilde{1}_{L} \rangle_{\scriptscriptstyle{\pm}\scriptscriptstyle{\pm}}\langle \tilde{1}_{L} \vert ) = \mathcal{L}_{\mathrm{cd}}(\vert \tilde{1}_{L} \rangle_{\scriptscriptstyle{\pm}\scriptscriptstyle{\pm}}\langle \tilde{1}_{L} \vert)  = 0$ independently of the parameters. Recall that $\mathcal{L}$ and $\mathcal{L}_{\mathrm{cd}}$ are defined in (\ref{EcMaestraFinalb}) and (\ref{Dephasing1}).  

In conclusion, $\vert 0_{L}\rangle$ and $\vert \tilde{1}_{L} \rangle_{\scriptscriptstyle{\pm}}$ are immune to all the processes embodied by $\mathcal{L}$, as well as to inhomogeneous broadening and correlated global and local dephasing described by $\mathcal{L}_{\mathrm{cd}}$ and $\mathcal{L}_{\mathrm{loc}}$. It is important to emphasize that this immunity is independent of the parameters appearing in all the Liouvillians (they are dark states). Therefore, $\vert 0_{L}\rangle$ and $\vert \tilde{1}_{L} \rangle_{\scriptscriptstyle{\pm}}$ could also be used as highly robust states of a logical qubit in the presence of local dephasing interactions. Unfortunately, both states are vulnerable to global and local uncorrelated dephasing.

The main conclusions obtained in this section for two qutrits can be extended to an arbitrary number of qutrits. In particular, the construction of the antisymmetric odd state (\ref{IHB4}) and mixed symmetry odd state (\ref{IHB5}) now translates to an odd subradiant state (pertaining to the direct sum of all representations except the symmetric, superradiant representation) and a mixed symmetry odd state, and follows the same lines that in the case of two qutrits. The only difference is that more freedom appears as the number of qutrits is increased.

\section{An origin of the model}
\label{Origen}

In this section we present one possible origin of the effective model presented in Sec.~\ref{SecQutrits}. 

\begin{figure}[b]
\includegraphics[scale=0.9]{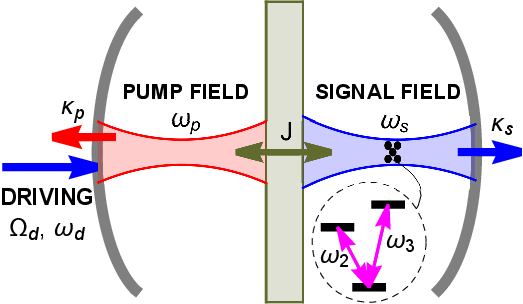}
\caption{\label{Figure2} The figure presents a schematic of the physical system under consideration. It is composed of a collection of $N\geq 2$ qutrits (represented by the black dots on the righthand side) and two single-mode cavity electromagnetic fields called the \textit{pump field} and the \textit{signal field}. The pump field has frequency $\omega_{p}$, is subject to dissipation at a rate $\kappa_{p}$, and is also driven by a classical field with frequency $\omega_{d}$ and driving strength $\Omega_{d}$. The signal field has frequency $\omega_{s}$, is subject to dissipation at a rate $\kappa_{s}$, and interacts with the pump field with coupling strength $J$. The qutrits have a V-configuration with transition frequencies $\omega_{j}$ and interact with the signal field with coupling strengths $g_{j}$ $(j=2,3)$, see also Fig.~\ref{Figure1}.}
\end{figure}

Consider a tripartite quantum system composed of $N\geq 2$ identical three-level atoms with a V-configuration and two single-mode cavity electromagnetic fields which will be called the \textit{pump} and \textit{signal} fields. The atoms, which will henceforth be called \textit{qutrits}, are bosons that only interact with the signal field, while the pump field is driven by a classical field and is parametrically coupled to the signal field. A schematic of the system is presented in Fig.~\ref{Figure2}. 

The Hamiltonian of the system is
\begin{eqnarray}
\label{Origen1}
H(t) &=& \sum_{l=p,s} \hbar \omega_{l} a_{l}^{\dagger}a_{l} + \sum_{j=2,3} \hbar \omega_{j}S_{jj} \cr
&&  + \hbar \Omega_{d} (a_{p}^{\dagger}e^{-i\omega_{d}t} + a_{p}e^{i\omega_{d}t}) + \hbar J \left[ a_{p} (a_{s}^{\dagger})^{2} + a_{p}^{\dagger}a_{s}^{2} \right]\cr
&& + \hbar \sum_{j=2,3} g_{j}(a_{s}^{\dagger} S_{1j} + a_{s} S_{1j}^{\dagger}) . 
\end{eqnarray}
Here $\omega_{l}>0$ is the angular frequency of the pump (signal) field if $l=p \ (s)$ and $a_{l}^{\dagger}$ and $a_{l}$ are the corresponding creation and annihilation operators. As presented in Sec.~\ref{SecQutrits}, the qutrits are described in second quantization and we have chosen the energy scale so that the ground level of the qutrits has energy equal to zero. The terms in the first line of the righthand side of (\ref{Origen1}) correspond to the free energies of all the component subsystems, while the terms in the second line describe the driving (in the rotating-wave-approximation) of the pump field and the parametric coupling between the two fields. The driving angular frequency $\omega_{d}>0$ is quasiresonant with the frequency of the pump field $\omega_{d} \simeq \omega_{p}$ and the driving strength $\Omega_{d}$ (a real constant) is weak $\vert \Omega_{d} \vert \ll \omega_{d}$. Meanwhile, the strength of the parametric coupling is described by the real quantity $J$. The third line of the righthand side of (\ref{Origen1}) describes the interaction (in the rotating-wave-approximation) between the qutrits and the signal field where $g_{j}$ is a real parameter representing the single-qutrit coupling strength with the signal field (it is the same for all the qutrits). Notice that we have not yet imposed the conditions that $\omega_{2} = \omega_{3}$ and that $g_{2} = g_{3}$. This will be done at the end. 

The set of Fock states $\{ \vert n_{l} \rangle : \ n\in \mathbb{Z}^{+} \}$ is an orthonormal basis for the state space of the pump (signal) field if $l=p \ (s)$, while the set $\beta_{q}$ of occupation number states in (\ref{betaa}) is an orthonormal basis for the state space of the $N$ qutrits. An orthonormal basis for the complete system is obtained by taking the tensor product of the previous three bases.

We assume that the pump $(l=p)$ and signal $(l=s)$ fields are coupled to independent thermal baths of harmonic oscillators at zero temperature (actually one only requires that each thermal bath has a temperature $T_{l}>0$ such that the expected number of thermal photons of frequency $\omega_{l}$ at that temperature is sufficiently small so that it can be neglected \cite{Carmichael1}). Then, we model the effect these thermal baths on the pump $(l=p)$ and signal $(l=s)$ fields by means of the dissipators $\kappa_{l}\mathcal{D}(a_{l})$ defined in (\ref{EcMaestra2}) with $\kappa_{l}>0$ the decay rates. The evolution of the complete system is then described by the GKLS master equation
\begin{eqnarray}
\label{Origen2}
\frac{d}{dt}\rho(t) &=& -\frac{i}{\hbar}[H(t) , \rho(t)] + \sum_{l=p,s}\kappa_{l}\mathcal{D}(a_{l})\rho(t) ,
\end{eqnarray}
with $\rho(t)$ the density operator of the complete system.

The idea of how one expects the system to behave is the following. First, $\omega_{d} \simeq \omega_{p}$ so the driving is quasiresonant with the pump field and, consequently, the pump field will approximately be driven to a coherent state where the expected valued of the number of photons is $>0$. Second, $\kappa_{s}, \omega_{s} \gg \omega_{p}, \omega_{2}, \omega_{3}$ so the signal field will be approximately in the vacuum state and the signal field and the ensemble of qutrits interact dispersively. The purpose of the signal field is to act as an intermediary between the pump field and the ensemble of qutrits. Third, $\omega_{p} \simeq 2 \omega_{2} \simeq 2 \omega_{3}$ so there is an effective interaction between the pump field and the qutrits mediated by the signal field and the processes where a pump photon is emitted or absorbed by two qutrits are favored, see \cite{NoriO}. In the following it is important to keep this intuitive picture in mind so that one recognizes how the successive approximations lead precisely to this behavior. In particular, the first adiabatic approximation will confirm that the signal field is approximately in the vacuum state and that it acts as an intermediary between the pump field and the qutrits, while the averaging will make explicit the aforementioned effective interaction between the pump field and the ensemble qutrits and the second adiabatic approximation will confirm that the pump field is approximately in a coherent state. 
      
To eliminate the time-dependence of the Hamiltonian, first pass to an interaction picture (IP) using the unitary transformation
 \begin{eqnarray}
 \label{Origen3}
 U_{I}(t) = e^{-\frac{i}{\hbar}H_{0}t} ,
 \end{eqnarray}
 where
 \begin{eqnarray}
 \label{Origen4}
 H_{0} &=& \hbar \omega_{d} a_{p}^{\dagger} a_{p} + \frac{\hbar \omega_{d}}{2} a_{s}^{\dagger} a_{s} - \frac{\hbar\omega_{d}}{2}S_{11} .
 \end{eqnarray}
 Then, the density operator in the IP is given by $\rho_{I} (t) = U_{I}^{\dagger} (t) \rho (t) U_{I} (t)$ and its evolution is described by the IP master equation
 \begin{eqnarray}
 \label{Origen5}
 \frac{d}{dt} \rho_{I}(t) &=& \mathcal{L}_{I} \rho_{I}(t) , \cr
 &=& -\frac{i}{\hbar} \left[ H_{I}, \rho_{I} (t) \right] + \sum_{l=p,s} \kappa_{l} \mathcal{D}(a_{l})\rho_{I}(t)  , \quad
 \end{eqnarray}
 where
 \begin{eqnarray}
 \label{Origen6}
 \frac{1}{\hbar}H_{I} &=& \delta_{s} a_{s}^{\dagger} a_{s} + \delta_{p} a_{p}^{\dagger} a_{p} + \sum_{j=2,3} \omega_{j} S_{jj} +  \frac{\omega_{d}}{2} S_{11} \cr
 && + \Omega_{d}( a_{p}^{\dagger} + a_{p} ) + J \left[ a_{p} (a_{s}^{\dagger})^{2} + a_{p}^{\dagger} a_{s}^{2} \right] \cr
 && + \sum_{j=2,3}  g_{j}(a_{s}^{\dagger}S_{1j} + a_{s}S_{1j}^{\dagger})  ,
 \end{eqnarray}
and we have introduced the detunings
 \begin{eqnarray}
 \label{Origen7}
 \delta_{p} &=& \omega_{p} - \omega_{d} , \quad\quad \delta_{s} = \omega_{s} - \frac{\omega_{d}}{2} .
 \end{eqnarray}
 We now perform the approximations.

\subsection{First adiabatic approximation}

The complete system can be decomposed in two subsystems: the \textit{signal field} on one hand, and the \textit{qutrits + driven pump field} on the other. 

Assume
\begin{eqnarray}
\label{Origen8}
\epsilon_{1} &=& \mathrm{max}\Bigg\{ \left\vert \frac{\delta_{p}}{\kappa_{s}} \right\vert, \  \frac{\omega_{j}}{\kappa_{s}} , \ \left\vert \frac{\Omega_{d}}{\kappa_{s}} \right\vert, \ \left\vert \frac{J}{\kappa_{s}} \right\vert, \ \left\vert \frac{g_{j}}{\kappa_{s}} \right\vert , \ \frac{\omega_{d}}{2\kappa_{s}}  , \ \frac{\kappa_{p}}{\kappa_{s}} :  \cr  
&& \quad\quad\quad\quad j=2,3  \Bigg\} \ \ll \ 1 . 
\end{eqnarray}
To preserve the order of the asymptotic expansion also assume that each of the nonzero elements of the set in (\ref{Origen8}) is $\gg \epsilon_{1}^{2}$ and that $\vert \delta_{s}/ \kappa_{s} \vert \gg \epsilon_{1}$. Observe that the quantities in the set (\ref{Origen8}) come from the coefficients of the terms appearing in the master equation (\ref{Origen5}) and (\ref{Origen6}) and dividing them by $\kappa_{s}$. 

The assumption implies that the signal field evolves rapidly and is strongly dissipative, while the qutrits+pump field is a subsystem that evolves slowly. Hence, one can adiabatically eliminate the signal field. In addition, measuring time in units of $1/\kappa_{s}$ (that is, expressing the master equation in (\ref{Origen5}) in terms of the dimensionless time $\tau = \kappa_{s}t$) leads to the conclusion that, to order zero in $\epsilon_{1}$, the signal field evolves according to the master equation of a damped harmonic oscillator at zero temperature: 
\begin{eqnarray}
\label{Origen9}
\frac{d}{dt} \rho_{I}(t) = -\frac{i}{\hbar} \left[ \hbar\delta_{s}a_{s}^{\dagger}a_{s} , \rho_{I}(t) \right] + \kappa_{s} \mathcal{D}(a_{s})\rho_{I}(t) .
\end{eqnarray}
As a consequence of (\ref{Origen9}), the signal field is a stable subsystem that converges to the stationary state $\vert 0_{s} \rangle \langle 0_{s} \vert$. 

Adiabatically eliminating the signal field from (\ref{Origen5}) \cite{Tesis,Tesis2}, to second order in $\epsilon_{1}$ the reduced dynamics of the \textit{qutrits+driven pump field} subsystem is given by
\begin{eqnarray}
\label{Origen10}
\frac{d}{dt} (\rho_{qp})_{I}(t) 
&=& -\frac{i}{\hbar} \left[ H_{qp} , \ (\rho_{qp})_{I}(t) \right] + \kappa_{p}' \mathcal{D}(a_{p})(\rho_{qp})_{I}(t) \cr
&&  + \kappa_{s} \mathcal{D}( S ) (\rho_{qp})_{I}(t),
\end{eqnarray}
where
\begin{eqnarray}
\label{Origen11}
\frac{1}{\hbar}H_{qp} 
&=&\delta_{p}' a_{p}^{\dagger}a_{p} +  \sum_{j=2,3} \omega_{j} S_{jj} +  \frac{\omega_{d}}{2} S_{11} - \delta_{s} S^{\dagger} S \cr
&&  + \Omega_{d} (a_{p}^{\dagger} + a_{p}) + \langle 0_{s} \vert  \frac{1}{\hbar}H_{\mathrm{int}} \vert 0_{s} \rangle , \cr
\frac{1}{\hbar}H_{\mathrm{int}} &=& J\left[ a_{p}(a_{s}^{\dagger})^{2} + a_{p}^{\dagger} a_{s}^{2} \right] + \kappa_{s}'(a_{s}^{\dagger}S + a_{s} S^{\dagger}) .
\end{eqnarray}
Here we have introduced the effective qutrit operator $S$ and the IP density operator $(\rho_{qp})_{I}(t)$ of the \textit{qutrits+driven pump field} subsystem by the equations
\begin{eqnarray}
\label{Origen12}
S &=& \frac{1}{\kappa_{s}'} \left( g_{2}S_{12} + g_{3}S_{13} \right), \cr
(\rho_{qp})_{I}(t) &=&  ( \mathrm{Tr}_{s}[\rho (t)] )_{I}(t).
\end{eqnarray}
We have also defined the effective parameters
\begin{eqnarray}
\label{Origen13}
\delta_{p}' &=& \delta_{p}  - \delta_{s}\left( \frac{J}{\kappa_{s}'} \right)^{2} , \quad \kappa_{p}' = \kappa_{p} + \kappa_{s}\left( \frac{J}{\kappa_{s}'} \right)^{2} , \cr
\kappa_{s} ' &=& \sqrt{\left( \frac{\kappa_{s}}{2} \right)^{2} + \delta_{s}^{2}} .
\end{eqnarray}
Notice that $\mathrm{Tr}_{s}$ is the partial trace with respect to the degrees of freedom of the signal field, so
\begin{eqnarray}
\label{Origen14}
(\rho_{qp})_{I}(t) &=& e^{-i(\omega_{d}t/2) S_{11}} e^{i(\omega_{d}t) a_{p}^{\dagger}a_{p}} \mathrm{Tr}_{s}[ \rho (t) ] \times \cr
&& e^{-i(\omega_{d}t) a_{p}^{\dagger}a_{p}} e^{i (\omega_{d}t/2) S_{11}}  .
\end{eqnarray}

Comparing (\ref{Origen5}) and (\ref{Origen6}) with (\ref{Origen10}) and (\ref{Origen11}) one finds that the effect of the interaction of the signal field with the \textit{qutrits+driven pump field} consists in an increase of the pump field decay rate and in the appearance of incoherent transitions from the excited states to the ground state in the qutrits, as described respectively by the increase of $\kappa_{p}$ to $\kappa_{p}'$ and by the dissipator $\kappa_{s} \mathcal{D}(S)$. It also includes a shift of the pump field frequency from $\delta_{p}$ to $\delta_{p}'$, a coherent interaction among qutrits described by $-\delta_{s}S^{\dagger}S$, and a coherent interaction between the pump field and the qutrits embodied by $\langle 0_{s} \vert  (1/\hbar)H_{\mathrm{int}} \vert 0_{s} \rangle$. In particular, observe that expanding $-\delta_{s}S^{\dagger}S$ using the definition of $S$ in (\ref{Origen12}) leads to 
\begin{eqnarray}
\label{Origen14b}
-\delta_{s}S^{\dagger}S
&=& -\frac{\delta_{s}}{(\kappa_{s}')^{2}}b_{1}b_{1}^{\dagger} \left[ \sum_{j=2,3} g_{j}^{2}S_{jj} + g_{2}g_{3}(S_{23} + S_{32}) \right] , \nonumber
\end{eqnarray}
so the interaction of the qutrits with the signal field also leads to shifts of the transitions frequencies $\omega_{j}$ $(j=2,3)$ and to an effective coupling between the excited levels of the qutrits, all of which depend on the number of particles in the ground state.

Observe that $\langle 0_{s} \vert H_{\mathrm{int}} \vert 0_{s} \rangle = 0$, so the pump field and the qutrits would evolve independently of one another according to (\ref{Origen10}) and (\ref{Origen11}). One would have to  preserve terms up to order $3$ in $\epsilon_{1}$ to include the effective interaction between the qutrits and the pump field induced by the signal field. This is rather difficult to calculate \cite{Tesis,Tesis2}, so we will take advantage of the different time-scales involved in the evolution of the system to average $H_{\mathrm{int}}$ and obtain an approximate Hamiltonian whose expected value in the state $\vert 0_{s} \rangle$ is not zero. This averaging has to be done to third order to be able to describe the effective qutrits-pump field interaction induced by the signal field. 

\subsection{Averaging}

Assume that
\begin{eqnarray}
\label{CondicionAvg}
\vert J \vert , \ \vert g_{2} \vert , \ \vert g_{3} \vert &\ll& \omega_{s} ,
\end{eqnarray}
and
\begin{eqnarray}
\label{James25b}
\omega_{s} \gg \omega_{p} > \omega_{j}  \quad (j=2,3).
\end{eqnarray}
Then, one can use \textit{James' averaging method} \cite{James1,James2} to third order [see Appendix \ref{AppendixC}] to obtain the approximation
\begin{eqnarray}
\label{Averaging}
&& \langle 0_{s} \vert \frac{1}{\hbar}H_{\mathrm{int}} \vert 0_{s} \rangle \cr
&\simeq& 
- (S_{11} + 1) \Bigg[ \frac{g_{2}^{2}}{\Delta_{2}}S_{22} + \frac{g_{3}^{2}}{\Delta_{3}}S_{33} +  g_{23} (S_{32} + S_{32}^{\dagger}) \Bigg]  \cr
&& -2\left( \frac{J^{2}}{\Delta_{s}} \right)a_{p}^{\dagger}a_{p} + \sum_{j=2,3} \chi_{j}\left[ S_{1j}^{2} a_{p}^{\dagger} + (S_{1j}^{\dagger})^{2} a_{p} \right] \cr
&& + \chi_{23}\left[ S_{12}S_{13} a_{p}^{\dagger} + S_{13}^{\dagger}S_{12}^{\dagger} a_{p} \right] ,
\end{eqnarray}
where 
\begin{eqnarray}
\label{Averaging2}
\chi_{j} &=& g_{j}^{2}J \left[ \frac{1}{2\Delta_{j}^{2}} + \frac{1}{\Delta_{s} (\Delta_{s} - \Delta_{j})} \right] , \quad (j=2,3), \cr
\chi_{23} &=& g_{2}g_{3} J \sum_{j=2,3} \left[ \frac{1}{\Delta_{j}(\Delta_{2} + \Delta_{3})} + \frac{1}{\Delta_{s}(\Delta_{s} - \Delta_{j})} \right] , \cr
g_{23} &=& \frac{g_{2}g_{3}}{2}\left( \frac{1}{\Delta_{2}} + \frac{1}{\Delta_{3}} \right) ,
\end{eqnarray}
and we have introduced the detunings
\begin{eqnarray}
\label{Origen15}
\Delta_{s} &=& 2\omega_{s} - \omega_{p}, \quad\quad \Delta_{j} = \omega_{s} - \omega_{j} \quad (j=2,3) . \quad
\end{eqnarray}
Then, $\langle 0_{s} \vert \frac{1}{\hbar}H_{\mathrm{int}} \vert 0_{s} \rangle$ is to be replaced by the righthand side of (\ref{Averaging}) in the effective master equation in (\ref{Origen10}) and (\ref{Origen11}). In the righthand side of (\ref{Averaging}) the terms multiplied by $\chi_{2}$, $\chi_{3}$, and $\chi_{23}$ are of order $3$, while the rest of the terms are of order $2$. Notice that the terms of order $3$ describe the effective interactions between the qutrits and the pump field where a pump photon is absorbed or emitted by two qutrits. Meanwhile, all the terms of order $2$ can be combined with terms already present in the master equation and simply lead to a modification of the values of $\delta_{p}'$ and $\delta_{s}/(\kappa_{s}')^{2}$. It is essential to perform the averaging method at least to order three to obtain the effective interaction between the qutrits and the pump field induced by the signal field.

\subsection{Second adiabatic approximation}

To obtain an effective master equation describing the dynamics of the $N$ qutrits we adiabatically eliminate the driven pump field.

Assume
\begin{eqnarray}
\label{Origen16}
\epsilon_{2} &=& \mathrm{max}\Bigg\{ \frac{\omega_{d}}{2\kappa_{p}'}  , \  \frac{\left\vert \omega_{j} - \frac{g_{j}^{2}}{\Delta_{j}} \right\vert}{\kappa_{p}'} , \ \frac{\vert \chi_{23} \vert}{\kappa_{p}'} , \ \frac{\vert \chi_{j} \vert }{\kappa_{p}'}  , \ \frac{\vert g_{23} \vert}{\kappa_{p}'} , \cr
&& \quad \left\vert \frac{g_{j}^{2}}{\Delta_{j} \kappa_{p}'} \right\vert , \ \frac{1}{\kappa_{p}'} \left\vert \frac{\delta_{s} g_{j}g_{k}}{(\kappa_{s}')^{2}} \right\vert : \ \ j,k=2,3 \Bigg\} \ll 1 .  \quad\quad
\end{eqnarray}
In order to preserve the order of the asymptotic expansion each nonzero element of the set in the definition of $\epsilon_{2}$ must be $\gg \epsilon_{2}^{2}$  and one must also have $\epsilon_{2} \ll \vert \Omega_{d} \vert / \kappa_{p}'$ and $\epsilon_{2} \ll \vert  \delta_{p}' - 2J^{2}/\Delta_{s} \vert / \kappa_{p}'$ (the last one if $\vert \delta_{p}' - 2J^{2}/\Delta_{s} \vert \not= 0$).

The assumption implies that the pump field evolves rapidly and is strongly dissipative, while the qutrits evolve slowly. Hence, one can adiabatically eliminate the pump field. In addition, measuring time in units of $1/\kappa_{p}'$ (that is, expressing the master equation in (\ref{Origen10}), (\ref{Origen11}), and (\ref{Averaging}) in terms of the dimensionless time $\tau = \kappa_{p}'t$) leads to the conclusion that, to order zero in $\epsilon_{2}$, the pump field evolves according to the master equation of a driven and damped harmonic oscillator at zero temperature: 
\begin{eqnarray}
\label{Origen17}
\frac{d}{dt} (\rho_{qp})_{I}(t) &=& -\frac{i}{\hbar} \left[ \hbar\delta_{p}'' a_{p}^{\dagger}a_{p} + \hbar\Omega_{d}(a_{p}^{\dagger} + a_{p}) , \ (\rho_{qp})_{I}(t) \right] \cr
&& + \kappa_{p}' \mathcal{D}(a_{p})(\rho_{qp})_{I}(t) ,
\end{eqnarray}
where the harmonic oscillator angular frequency is
\begin{eqnarray}
\label{Origen18}
\delta_{p}'' &=& \delta_{p}' -2\frac{J^{2}}{\Delta_{s}} .
\end{eqnarray}
As a consequence of (\ref{Origen17}), the pump field is a stable subsystem that converges to the stationary state $\vert \alpha_{p} \rangle \langle \alpha_{p} \vert$. Here $\vert \alpha_{p} \rangle$ is a coherent state with
\begin{eqnarray}
\label{Origen19}
\alpha_{p} &=& \frac{\Omega_{d}}{ - \delta_{p}''  + i \left( \frac{\kappa_{p}'}{2} \right) } .
\end{eqnarray} 

Adiabatically eliminating the pump field from the master equation in (\ref{Origen10}), (\ref{Origen11}), and (\ref{Averaging}) \cite{Tesis,Tesis2}, to second order in $\epsilon_{2}$ the reduced dynamics of the ensemble of $N$ qutrits is given by
\begin{eqnarray}
\label{Origen20}
\frac{d}{dt} (\rho_{q})_{I}(t) 
&=& -\frac{i}{\hbar} \left[ H_{\mathrm{eff}}, \ (\rho_{q})_{I}(t) \right] + \kappa_{s}\mathcal{D}(S)(\rho_{q})_{I}(t) \cr
&& \quad\quad\quad + \kappa_{p}' \mathcal{D}(S_{q})(\rho_{q})_{I}(t) ,
\end{eqnarray}
where the effective Hamiltonian $H_{\mathrm{eff}}$ of the qutrits is
\begin{eqnarray}
\label{Origen21}
\frac{1}{\hbar}H_{\mathrm{eff}} 
&=& \frac{\omega_{d}}{2} S_{11} + \sum_{j=2,3} \left[ \xi_{j} + (\xi_{j} - \omega_{j})S_{11} \right] S_{jj} \cr
&& - \xi_{23}(S_{11} + 1) (S_{32} + S_{32}^{\dagger}) \cr
&& - \delta_{p}'' S_{q}^{\dagger} S_{q} +  \left( \alpha_{q}^{*}S_{q} + \alpha_{q} S_{q}^{\dagger} \right) .
\end{eqnarray}
Here we have defined the parameters
\begin{eqnarray}
\label{EcMaestra3p3}
\xi_{j} &=& \omega_{j} - g_{j}^{2} \left[ \frac{1}{\Delta_{j}} + \frac{\delta_{s}}{(\kappa_{s}')^{2}} \right] , \quad (j=2,3), \cr
\xi_{23} &=& g_{2}g_{3} \left[ \frac{1}{2}\left( \frac{1}{\Delta_{2}} + \frac{1}{\Delta_{3}} \right) + \frac{\delta_{s}}{(\kappa_{s}')^{2}}\right] , \cr
\alpha_{q} &=& \alpha_{p} \kappa_{p}'' ,
\end{eqnarray}
and we have introduced the qutrit operator
\begin{eqnarray}
\label{EcMaestra3p2}
S_{q} &=& \frac{1}{\kappa_{p}''}\left( \chi_{2}S_{12}^{2} + \chi_{3}S_{13}^{2} + \chi_{23} S_{12} S_{13} \right) ,
\end{eqnarray}
with the pump field version of $\kappa_{s}'$ [see (\ref{Origen13})]
\begin{eqnarray}
\label{Origen23}
\kappa_{p}'' &=& \sqrt{\left( \frac{\kappa_{p}'}{2} \right)^{2} + (\delta_{p}'')^{2}} .
\end{eqnarray}
Also, $(\rho_{q})_{I}(t)$ is the density operator of the qutrits in the IP. Explicitly,
\begin{eqnarray}
\label{EcMaestra3p4}
(\rho_{q})_{I}(t) &=& e^{-i(\omega_{d}t/2) S_{11}} \mathrm{Tr}_{p+s} [\rho(t)] e^{i(\omega_{d}t/2) S_{11}} , \quad\quad
\end{eqnarray}
where $\mathrm{Tr}_{s+p}$ denotes the partial trace with respect to the degrees of freedom of the pump and signal fields.

Comparing (\ref{Origen10}) and (\ref{Origen11}) with (\ref{Origen20}) and (\ref{Origen21})  one finds that the interaction between the qutrits and the pump field induces a quadratic dissipator $\kappa_{p}'\mathcal{D}(S_{q})$ for the qutrits, as well as shifts for $\omega_{2}$, $\omega_{3}$, and the strength of the excited levels coupling in the second line of (\ref{Origen21}). Also notice the appearance of the coherent interactions in the last line of (\ref{Origen21}): the term $(\alpha_{q}^{*}S_{q} + \alpha_{q}S_{q}^{\dagger})$ is similar to a squeezing operator for $b_{j}$ and $b_{j}^{\dagger}$ $(j=2,3)$, while $-\delta_{p}''S_{q}^{\dagger}S_{q}$ has fourth order terms involving $b_{2}$, $b_{3}$, and their adjoints.  Finally, observe that the linear dissipator $\kappa_{s}\mathcal{D}(S)$ is not affected.

\subsection{The final model}

Now assume that 
\begin{eqnarray}
\label{Final}
g_{2} &=& g_{3} = g, \quad \omega_{2} = \omega_{3} = \omega_{q} ,
\end{eqnarray}
and consider the operator $S_{-}$ defined in (\ref{Sij}).

Define the quantities
\begin{eqnarray}
\label{Final2}
\kappa_{1} &=& \frac{\kappa_{s} g^{2}}{(\kappa_{s}')^{2}} , \quad\quad \Delta = \omega_{s} - \omega_{q} , \quad\quad \delta = - \delta_{p}'' \left( \frac{\kappa_{2}}{\kappa_{p}'} \right) ,  \cr 
\kappa_{2} &=& \frac{\kappa_{p}' \chi^{2}}{(\kappa_{p}'')^{2}} , \quad\quad \delta_{1} = \frac{\omega_{d}}{2} - \omega_{q}, \quad \ \ \xi = g^{2} \left[  \frac{1}{\Delta} + \frac{\delta_{s}}{(\kappa_{s}')^{2}} \right]  , \cr
\alpha_{0} &=& \alpha_{p} \chi  , \quad\quad \ \chi = g^{2} J \left[ \frac{1}{2\Delta^{2}} + \frac{1}{\Delta_{s}(\Delta_{s}- \Delta)} \right] .
\end{eqnarray}

Using the assumption in (\ref{Final}) and the quantities in (\ref{Final2}) if follows that 
\begin{eqnarray}
\label{Final3}
S &=& \sqrt{\frac{\kappa_{1}}{\kappa_{s}}} S_{-} , \quad\quad  \Delta _{2} = \Delta_{3} = \Delta , \quad\quad \alpha_{q} = \alpha_{0} \sqrt{\frac{\kappa_{p}'}{\kappa_{2}}} , \cr
S_{q} &=&\sqrt{\frac{\kappa_{2}}{\kappa_{p}'}} S_{-}^{2} , \quad\quad \chi_{2} = \chi_{3} = \chi , \quad\quad \ \   \chi_{23} = 2\chi , \cr 
\xi_{23} &=& \xi , \quad\quad\quad\quad\quad \xi_{2} = \xi_{3} = \omega_{q} - \xi  .
\end{eqnarray}
If one uses (\ref{Final2}), (\ref{Final3}), and $(S_{11} + S_{22} + S_{33}) = N$ because there are $N$ qutrits, then the master equation in (\ref{Origen20})-(\ref{Origen21}) takes the form of the master equation in (\ref{EcMaestraFinal})-(\ref{EcMaestraFinal2}). Note that the density operator $\rho(t)$ appearing in (\ref{EcMaestraFinal}) is actually $(\rho_{q})_{I}(t)$, which is defined in (\ref{EcMaestra3p4}). In particular, the detuning $\delta_{1}$ defined in (\ref{Final2}) can be tuned to zero or to a positive or negative value because one can adjust the frequency $\omega_{d}$ of the classical field driving the pump field. 

\subsection{The case of qubits}

We now discuss what happens if one has qubits instead of qutrits. Moreover, it allows one to compare the master equation in (\ref{EcMaestraFinal})-(\ref{EcMaestraFinal2}) with that deduced in \cite{NoriO}.

To obtain the case of qubits one simply has to take
\begin{eqnarray}
\label{qubit1}
\omega_{3} &=& g_{3} = 0, \quad \omega_{q} = \omega_{2}, \quad g = g_{2}, \quad \chi = \chi_{2}. \quad
\end{eqnarray} 
in the original Hamiltonian in (\ref{Origen1}). Then, one performs the same approximations as before to obtain the effective $N$ qubit master equation
\begin{eqnarray}
\label{qubit2}
\frac{d}{dt}(\rho_{q})_{I}(t) &=& -\frac{i}{\hbar}[ H_{\mathrm{eff}}, \ (\rho_{q})_{I}(t) ] + \kappa_{1} \mathcal{D}(S_{12}) (\rho_{q})_{I}(t) \cr
&& \quad + \kappa_{2} \mathcal{D}(S_{12}^{2}) (\rho_{q})_{I}(t),
\end{eqnarray}
with the quantities defined in (\ref{Final2}) and (\ref{qubit1}) and
\begin{eqnarray}
\label{qubit3}
\frac{1}{\hbar} H_{\mathrm{eff}} &=& \frac{\omega_{d}}{2}S_{11} + \left( \omega_{q} - \xi - \xi S_{11} \right) S_{22}  \cr
&& + \delta (S_{12}^{2})^{\dagger} S_{12}^{2} + \alpha_{0}^{*} S_{12}^{2} + \alpha_{0} (S_{12}^{2} )^{\dagger} . \quad\quad
\end{eqnarray}
This master equation was obtained from (\ref{Origen20})-(\ref{EcMaestra3p2}) by applying (\ref{qubit1}). In particular, notice that (\ref{qubit1}) leads to
\begin{eqnarray}
\xi_{3} &=& \chi_{3} = 0 , \quad\quad\quad\quad\quad  S = \frac{g}{\kappa_{s}'} S_{12} , \cr
\xi_{23} &=& \chi_{23} = g_{23} = 0, \quad\quad S_{q} = \frac{\chi}{\kappa_{p}''}S_{12}^{2} . \nonumber
\end{eqnarray}

We now reexpress $H_{\mathrm{eff}}$. Using the commutation relations of the $b_{j}$ and their adjoints and the fact that there are $N$ qubits (so $S_{11} + S_{22} = N$), one has 
\begin{eqnarray}
\frac{\omega_{d}}{2} S_{11} + (\omega_{q} - \xi - \xi S_{11})S_{22} = \omega_{q} N + \delta_{1} S_{11} - \xi S_{12}^{\dagger} S_{12} . \nonumber
\end{eqnarray}
Substituting this result in (\ref{qubit3}) leads to
\begin{eqnarray}
\label{qubit4}
\frac{1}{\hbar} H_{\mathrm{eff}} &=& \delta_{1} S_{11} - \xi S_{12}^{\dagger} S_{12} +\delta (S_{12}^{2})^{\dagger} S_{12}^{2}  \cr
&& + \alpha_{0}^{*} S_{12}^{2} + \alpha_{0} (S_{12}^{2} )^{\dagger} , \quad\quad
\end{eqnarray}
where we have discarded the term $\omega_{q} N$ because it is a scalar that disappears when one substitutes $H_{\mathrm{eff}}$ in the commutator of the master equation in (\ref{qubit2}).

Observe that the $N$ qubit master equation in (\ref{qubit2}) and (\ref{qubit4}) has exactly the same form as the $N$ qutrit master equation in (\ref{EcMaestraFinal})-(\ref{EcMaestraFinal2}) because $S_{-} = S_{12}$ for qubits (there is no third level). 

We now compare the $N$ qubit master equation in (\ref{qubit2}) and (\ref{qubit4}) with the one deduced in \cite{NoriO}. The $N$ qubit master equation in \cite{NoriO} has exactly the same dissipators and the same driving in the Hamiltonian but with $\kappa_{1at}$, $\kappa_{2at}$, and $i\chi_{2at}/N$ replacing our $\kappa_{1}$, $\kappa_{2}$, and $\alpha_{0}^{*}$, respectively. Ref. \cite{NoriO} considers $\kappa_{p}' \simeq \kappa_{p}$; $\vert \delta_{s} \vert \gg (\kappa_{s}/2)$; $\kappa_{p}'/2 \gg \vert \delta_{p}'' \vert$; $\kappa_{p}'/2 \gg \vert 2J^{2}/\Delta_{s} - \delta_{p}' \vert$; and $\omega_{d} = \omega_{p} = 2 \omega_{q}$. Using these conditions one finds that $\kappa_{2} = \kappa_{2at}$, $\kappa_{1} = \kappa_{1at}$, $\alpha_{0}^{*} = i\chi_{2at}/N$, and $\delta_{1} = 0$, so the two master equations are identical under these conditions except that \cite{NoriO} neglects the last two terms in the first line of the righthand side of (\ref{qubit4}).

\subsection{Approximate stationary solutions of the original master equation}

Sec.~\ref{solEst} presented two parameter independent stationary states, $\vert 0_{L}\rangle\langle 0_{L} \vert$ and $\vert 1_{L}\rangle\langle 1_{L} \vert$, of the effective master equation in (\ref{EcMaestraFinal}) and (\ref{EcMaestraFinalb}) that can be used as logical states of a qubit. In this section we show that they are part of an approximate stationary solution of the original master equation in (\ref{Origen5}) and (\ref{Origen6}) when (\ref{Final}) and the assumptions of the first (\ref{Origen8}) and second (\ref{Origen16}) adiabatic approximations hold [see also the paragraphs below (\ref{Origen8}) and (\ref{Origen16})].

Using (\ref{Final}) and $N = S_{11} + S_{22} + S_{33}$ one can write the Hamiltonian in (\ref{Origen6}) as
\begin{eqnarray}
\label{ExtraF1}
\frac{1}{\hbar} H_{I} &=&
\delta_{s} a_{s}^{\dagger}a_{s} + \delta_{p}a_{p}^{\dagger}a_{p} + \omega_{q}(N-S_{11}) + \frac{\omega_{d}}{2}S_{11} \cr
&& + \Omega_{d}(a_{p}^{\dagger} + a_{p}) + J \left[ a_{p}(a_{s}^{\dagger})^{2} + a_{p}^{\dagger} a_{s}^{2} \right] \cr
&& + g (a_{s}^{\dagger}S_{-} + a_{s}S_{-}^{\dagger}) .
\end{eqnarray}

Now consider the density operators $(\mathbb{j} = 0,1)$
\begin{eqnarray}
\label{ExtraF2}
\rho_{\scriptscriptstyle{\mathbb{j}}} &=& \vert 0_{s}\rangle\langle 0_{s} \vert \otimes \vert \alpha_{po} \rangle\langle \alpha_{po} \vert \otimes \vert \mathbb{j}_{L} \rangle\langle \mathbb{j}_{L} \vert , \quad
\end{eqnarray}
where
\begin{eqnarray}
\label{ExtraF3}
\alpha_{po} &=& \frac{\Omega_{d}}{-\delta_{p} + i \left( \frac{\kappa_{p}}{2} \right)} \ .
\end{eqnarray}
It is important to note that $\vert 0_{s}\rangle\langle 0_{s} \vert$ is the steady-state solution of the damped harmonic oscillator master equation at zero temperature in (\ref{Origen9}), while $\vert \alpha_{po} \rangle\langle \alpha_{po} \vert$ is the steady-state solution of the driven and damped harmonic oscillator master equation at zero temperature in (\ref{Origen17}) with $\delta_{p}$ and $\kappa_{p}$ replacing $\delta_{p}''$ and $\kappa_{p}'$, respectively. Notice that we use $\alpha_{po}$ instead of $\alpha_{p}$ in (\ref{Origen19}) because it is $\delta_{p}$ and $\kappa_{p}$ that appear in the original master equation (\ref{Origen5}) and (\ref{ExtraF1}) instead of $\delta_{p}''$ and $\kappa_{p}'$. Using these observations and the assumptions of the first and second adiabatic approximations one can show [see Appendix \ref{AppF}] that for $n=1,2,...$ and $\mathbb{j} = 0,1$ one has
\begin{eqnarray}
\label{ExtraF5}
\mathcal{L}_{I}^{n} \rho_{\scriptscriptstyle{\mathbb{j}}} 
&\simeq&
-i\sqrt{2} J \Bigg[
\alpha_{po} (-k_{s} - i2\delta_{s})^{n-1} \vert 2_{s} \rangle\langle 0_{s} \vert \cr
&& \quad -\alpha_{po}^{*} (-\kappa_{s} + i2\delta_{s} )^{n-1}  \vert 0_{s} \rangle\langle 2_{s} \vert \Bigg]  \otimes \cr
&& \quad\quad \otimes \vert \alpha_{po} \rangle\langle \alpha_{po} \vert \otimes \vert \mathbb{j}_{L} \rangle\langle \mathbb{j}_{L} \vert . \quad
\end{eqnarray}
Since the interaction and the Schr\"{o}dinger pictures coincide at time $t=0$ [see (\ref{Origen3})], using (\ref{ExtraF5}) one finds that, according to the original master equation, $\rho_{\scriptscriptstyle{\mathbb{j}}}$ evolves as
\begin{eqnarray}
\label{ExtraF6}
\rho_{\scriptscriptstyle{\mathbb{j}}}^{I} (t) &=& e^{\mathcal{L}_{I}t}\rho_{\scriptscriptstyle{\mathbb{j}}} = \rho_{\scriptscriptstyle{\mathbb{j}}} + \sum_{n=1}^{\infty} \frac{t^{n}}{n!}\mathcal{L}_{I}^{n} \rho_{\scriptscriptstyle{\mathbb{j}}} , \cr
&\simeq& \rho_{\scriptscriptstyle{\mathbb{j}}} + R_{s}(t) \otimes \vert \alpha_{po} \rangle\langle \alpha_{po} \vert \otimes \vert \mathbb{j}_{L} \rangle\langle \mathbb{j}_{L} \vert ,
\end{eqnarray}
where
\begin{eqnarray}
\label{ExtraF7}
R_{s}(t) &=& i \frac{\sqrt{2} J \alpha_{po}}{\kappa_{s} + i2\delta_{s}} \left[ e^{-(\kappa_{s} + i 2\delta_{s})t} -1 \right] \vert 2_{s} \rangle\langle 0_{s} \vert + \mathrm{h.c.}. \quad\quad
\end{eqnarray}
Here h.c. stands for \textit{hermitian conjugate}.

Observe that 
\begin{eqnarray}
\label{ExtraF8}
\lim_{t\rightarrow + \infty} R_{s}(t) &=& -i \frac{\sqrt{2} J \alpha_{po}}{\kappa_{s} + i2\delta_{s}} \vert 2_{s} \rangle\langle 0_{s} \vert + \mathrm{h.c.}, \quad\quad
\end{eqnarray}
and, using the definition of $\alpha_{po}$ in (\ref{ExtraF3}), that 
\begin{eqnarray}
\label{ExtraF9}
\left\vert -i \frac{\sqrt{2} J \alpha_{po}}{\kappa_{s} + i2\delta_{s}} \right\vert &\leq&  2\sqrt{2} \epsilon_{1} \frac{\vert \Omega_{d} \vert}{\kappa_{p}} .
\end{eqnarray}
By the first and second adiabatic approximations the righthand side of (\ref{ExtraF9}) is $\ll 1$, so one concludes from (\ref{ExtraF6})-(\ref{ExtraF9}) that $\rho_{\scriptscriptstyle{\mathbb{j}}}^{I}(t)$ approximately tends to $\rho_{\scriptscriptstyle{\mathbb{j}}}$ as $t\rightarrow + \infty$.
 
\section{Conclusions}

We have presented a special case of a cat code where a logical qubit is encoded using engineered even and odd Schr\"{o}dinger cat states of an ensemble of identical qutrits with a symmetrical V-configuration (\textit{a qubit-disguised qutrit}). These logical states have either zero or one qutrits in the ground level and constitute dark states. In particular, they are immune to single-qutrit decay, two-qutrit decay and driving processes, and collective correlated dephasing. It is important to emphasize that these logical states do not depend on the parameters of the effective master equation describing the evolution of the ensemble of qutrits. In the case of two qutrits, the logical state encoded in the even cat state is immune to both inhomogeneous broadening and to local correlated dephasing, while the logical state encoded in the odd cat state is vulnerable to both processes. Nevertheless, in the presence of these permutation symmetry breaking interactions one can replace the fragile logical state with one with mixed permutation symmetry that is immune to both inhomogeneous broadening and local correlated dephasing and also to all the interactions to which the fragile logical state is invulnerable. The results obtained for two qutrits can be extrapolated to any number of qutrits with suitable modifications.

The effective master equation describing the evolution of the ensemble of qutrits can be deduced from the master equation describing a physical system composed of two parametrically coupled cavity-fields where one is driven by a classical field and the other interacts dispersively with an ensemble of three-level atoms that play the role of the qutrits. Both cavity-fields are subject to decay by the single-photon loss process and are assumed to be strongly dissipative so that they can be adiabatically eliminated. It is shown analytically that the logical qubit is part of an approximate stationary state of the physical system under the assumptions made to derive the effective model.

In principle this physical system can be implemented experimentally using an extension of the setup originally proposed in \cite{squeezing}: one considers two state-of-the-art superconducting coplanar waveguide resonators (CPWR), a superconducting quantum interference device (SQUID), and an ensemble of alkali atoms placed close to one of the CPWR. The SQUID is in charge of parametric coupling between the fields in each CPWR and, in particular, \cite{RydbergQubit} already realized the dispersive coupling between an ensemble of rubidium 87 atoms and a CPWR. The CPWR in \cite{RydbergQubit} naturally couples three hyperfine ground state levels of rubidium 87 in a V-configuration: $\vert F=2,M_{F}=-2\rangle$ and $\vert F=2,M_{F}=0\rangle$ play the role of the excited levels, while $\vert F=1,M_{F}=-1\rangle$ plays the role of the ground level. Although the aforementioned excited levels have different frequencies in the setup of \cite{RydbergQubit}, this difference could be controlled by decreasing the magnitude of the  magnetic field responsible for this detuning.

In future work we will investigate the effect of a detuning between the excited state frequencies (i.e. asymmetrical V-configuration) on the stationary states of the effective qutrit master equation, as well as the possibility to encode a logical qubit using these stationary states. In particular, we shall study the robustness of such states under dissipation and dephasing processes, as well as the interaction of two ensembles of qutrits (each one corresponding to a logical qubit) in order to define two-qubit logical gates. We also aim at investigating the advantages and disadvantages offered by ensembles of qutrits with a $\Lambda-$ or $\Sigma-$configuration. The inclusion of a thermal bath with $T>0$ for the signal and pump fields and how this will affect the robustnest of the logical qubit here introduced will also be addressed.

\section*{Acknowledgments}
We thank the support of the Spanish MICINN through
the project PID2022-138144NB-I00. L.O. Casta\~nos-Cervantes thanks the University of Ja\'en for its hospitality.


\appendix

\section{Parameter independent stationary solutions}
\label{AppendixA}

In this appendix we first prove three propositions that are then used to calculate the parameter independent stationary states of the master equation in (\ref{EcMaestraFinal}).

We first prove the following proposition:
\begin{eqnarray} 
\label{Prueba1A}
\mbox{$\lambda$ is an eigenvalue of $S_{-}$} \quad &\Rightarrow& \quad \lambda = 0.
\end{eqnarray} 
Let $\lambda$ be a nonzero complex number. Assume that $\lambda$ is an eigenvalue of $S_{-}$. Then there is a nonzero ket $\vert \Phi \rangle$ such that $S_{-}\vert \Phi \rangle = \lambda \vert \Phi \rangle$.

Since $\beta_{q}$ is an orthonormal basis for the state space of the $N$ qutrits, one has the closure relation 
\begin{eqnarray}
\label{Prueba1}
&& \sum_{n_{3}=0}^{N} \sum_{n_{2}=0}^{N-n_{3}} \vert N - n_{2} - n_{3}, n_{2}, n_{3} \rangle\langle N-n_{2} -n_{3}, n_{2}, n_{3} \vert \cr
&& = \ \mathbb{I}_{q} ,
\end{eqnarray}
where $\mathbb{I}_{q}$ is the identity operator in the state space of the $N$ qutrits. Introducing this closure relation between $S_{-}$ and $\vert \Phi \rangle$ it is straightforward to show that
\begin{eqnarray}
\label{Prueba2}
S_{-} \vert \Phi \rangle &=& \sum_{n_{3} = 0}^{N-1} \sum_{n_{2} = 0}^{N-n_{3}-1} \sqrt{N-n_{2}-n_{3}} \Big[ c_{n_{2}+1,n_{3}}\sqrt{n_{2}+1}  \cr
&&  + c_{n_{2},n_{3}+1} \sqrt{n_{3}+1} \Big] \vert N-n_{2}-n_{3}, n_{2}, n_{3} \rangle , \quad
\end{eqnarray}
where
\begin{eqnarray}
\label{Prueba3}
c_{n_{2},n_{3}} &=& \langle N -n_{2}-n_{3}, n_{2},n_{3} \vert \Phi \rangle .
\end{eqnarray}
Using the expansions of $S_{-}\vert \Phi \rangle$ and $\vert \Phi \rangle$ with respect to the basis $\beta_{q}$ it follows that
\begin{eqnarray}
\label{Prueba5}
&& S_{-}\vert \Phi \rangle = \lambda \vert \Phi \rangle \cr
&\Leftrightarrow&
\left\{ 
\begin{array}{c}
c_{0,N} = 0, \cr
c_{N-m_{3},m_{3}} = 0 \ \ \ (m_{3}=0,1,...,N-1), \cr
\lambda c_{n_{2},n_{3}} = \sqrt{N-n_{2}-n_{3}} \Big[ c_{n_{2}+1,n_{3}} \sqrt{n_{2}+1} \cr
 \quad \quad \quad + c_{n_{2},n_{3}+1} \sqrt{n_{3}+1} \Big] \cr
(n_{2} = 0,1,...N-n_{3}-1; \ n_{3} = 0,1,...,N-1).
\end{array}
\right. \cr
&&
\end{eqnarray}
It is clear that all the conditions on the righthand side of (\ref{Prueba5}) are satisfied if $c_{n_{2},n_{3}} = 0$ for all $n_{2} = 0, ..., N-n_{3}$ and $n_{3} = 0, ..., N$. We now prove that these conditions imply that $c_{n_{2},n_{3}} = 0$ for all $n_{2} = 0, ..., N-n_{3}$ and $n_{3} = 0, ..., N$.

Assume that all the conditions on the righthand side of (\ref{Prueba5}) hold. We claim that $c_{n_{2},n_{3}} = 0$ for all $n_{2} = 0,1, ..., N-n_{3}-1$ and $n_{3} = 0,1,...,N-1$. 

We prove the claim by induction over $\mu_{3}$ with $n_{3} = N-1-\mu_{3}$ for all $\mu_{3} = 0,1,...,N-1$. The induction parameter is $\mu_{3}$ instead of $n_{3}$ so one can start with $\mu_{3} = 0$ and $n_{3} = N-1$ instead of $n_{3} = 0$.

First assume that $\mu_{3} = 0$. Then $n_{3} = N-1$ and $n_{2} = 0$. From the third condition in (\ref{Prueba5}) one has
\begin{eqnarray}
\label{Prueba6}
\lambda c_{0,N-1} = c_{1,N-1} + c_{0,N} \sqrt{N}.
\end{eqnarray}
Using the first and the second conditions in (\ref{Prueba5}) one finds that the righthand side of (\ref{Prueba6}) is zero. Since $\lambda \not= 0$, it follows that $c_{0,N-1} = 0$ and the claim holds for $\mu_{3} = 0$.

Now assume that the claim holds for some $\mu_{3} \in \{ 0,..., N-2 \}$. We prove that it holds for $\mu_{3}+1$. In this case $n_{3} = N-2-\mu_{3}$ and $n_{2} = 0,1, ..., \mu_{3}+1$. 

From the third condition in (\ref{Prueba5}) and for $n_{2} = \mu_{3}+1$
\begin{eqnarray}
\label{Prueba7}
\lambda c_{\mu_{3}+1, N-2-\mu_{3}} &=& c_{\mu_{3}+2,N-2-\mu_{3}} \sqrt{\mu_{3}+2} \cr
&& \quad + c_{\mu_{3}+1,N-1-\mu_{3}}\sqrt{\mu_{3}+1} .
\end{eqnarray} 
Using the first and the second conditions in (\ref{Prueba5}) one finds that the righthand side of (\ref{Prueba7}) is zero, so 
\begin{eqnarray}
\label{Prueba7b}
c_{\mu_{3}+1,N-2-\mu_{3}} &=& 0.
\end{eqnarray}
Applying the induction hypothesis to the third condition in (\ref{Prueba5}) for $n_{2} = 0,1,..., \mu_{3}$ one obtains
\begin{eqnarray}
\label{Prueba8}
\lambda c_{\mu_{3},N-2-\mu_{3}} &=&  c_{\mu_{3}+1,N-2-\mu_{3}}\sqrt{2(\mu_{3}+1)} , \cr
\lambda c_{\mu_{3}-1,N-2-\mu_{3}} &=&  c_{\mu_{3},N-2-\mu_{3}} \sqrt{3\mu_{3}}, \cr
&\vdots& \cr
\lambda c_{0,N-2-\mu_{3}} &=&  c_{1,N-2-\mu_{3}} \sqrt{(\mu_{3}+2)}.
\end{eqnarray}
Substituting (\ref{Prueba7b}) in the righthand side of the first equation in (\ref{Prueba8}) one finds that $c_{\mu_{3},N-2-\mu_{3}} = 0$. Substituting this result in the second equation of (\ref{Prueba8}) one obtains that $c_{\mu_{3}-1,N-2-\mu_{3}} = 0$. Continuing in this manner one concludes that $c_{\mu_{3}-2,N-2-\mu_{3}} = 0$, ..., $c_{0,N-2-\mu_{3}} = 0$. Therefore, the claim holds for $\mu_{3}+1$. Consequently, the claim holds for all $\mu_{3} = 0,1,...,N-1$.

Using the claim it follows that the conditions in (\ref{Prueba5}) are equivalent to $c_{n_{2},n_{3}} = 0$ for all $n_{2}=0,1,...,N-n_{3}$ and $n_{3} = 0,1,...,N$, which in turn is equivalent to $\vert \Phi \rangle = 0$. But $\vert \Phi \rangle \not= 0$ because it is an eigenvector with $\lambda \not= 0$. Hence, we arrive at a contradiction that stems from the assumption that $\lambda$ is a nonzero eigenvalue of $S_{-}$. Therefore, we conclude that $\lambda$ is not an eigenvalue of $S_{-}$ if $\lambda \not= 0$. Hence, the proposition in (\ref{Prueba1A}) is true.

Consider a pure state $\vert\Phi\rangle\langle\Phi\vert$. We now prove that
\begin{eqnarray}
\label{Prueba2A}
\mathcal{D}(S_{-})\vert \Phi \rangle\langle \Phi \vert =0 \Leftrightarrow S_{-}\vert \Phi \rangle = 0.
\end{eqnarray}

Inspecting the form of $\mathcal{D}(S_{-})\vert\Phi\rangle\langle\Phi\vert$, it is clear that $\mathcal{D}(S_{-})\vert \Phi \rangle\langle \Phi \vert =0$ if $S_{-}\vert \Phi \rangle = 0$.

Now assume that $\mathcal{D}(S_{-})\vert\Phi\rangle\langle\Phi\vert = 0$. Since $\vert \Phi \rangle$ is normalized, it follows that
\begin{eqnarray}
\label{Proof1}
0 &=& \langle \Phi \vert \left[  \ \mathcal{D}(S_{-}) \vert \Phi \rangle\langle \Phi \vert \ \right]  \vert \Phi \rangle \cr
&=&  \langle \Phi \vert \left[ S_{-} \vert \Phi \rangle\langle \Phi \vert S_{-}^{\dagger} - \frac{1}{2}\left\{ S_{-}^{\dagger}S_{-},  \vert \Phi \rangle\langle \Phi \vert \right\} \right] \vert \Phi \rangle , \cr
&=& \langle \Phi \vert S_{-}\vert \Phi \rangle\langle \Phi \vert S_{-}^{\dagger} \vert \Phi \rangle - \frac{1}{2} \langle \Phi \vert S_{-}^{\dagger}S_{-}\vert \Phi \rangle - \frac{1}{2}\langle \Phi \vert S_{-}^{\dagger}S_{-} \vert \Phi \rangle , \cr
&=& \vert \langle \Phi \vert S_{-}\vert \Phi \rangle \vert^{2}  - \langle \Phi \vert S_{-}^{\dagger}S_{-}\vert \Phi \rangle  , \cr
&=& \vert \langle \Phi \vert S_{-}\Phi \rangle \vert^{2}  - \langle S_{-} \Phi \vert S_{-}\Phi \rangle  , 
\end{eqnarray}
so
\begin{eqnarray}
\label{Proof2}
\vert \langle \Phi \vert S_{-}\Phi \rangle \vert^{2}  = \langle S_{-} \Phi \vert S_{-}\Phi \rangle  .
\end{eqnarray}
From the Cauchy-Schwarz inequality one has
\begin{eqnarray}
\label{Proof3}
\vert \langle \Phi \vert S_{-}\Phi \rangle \vert^{2}  \leq \langle S_{-} \Phi \vert S_{-}\Phi \rangle  .
\end{eqnarray}
From (\ref{Proof2}) it follows that the equality in (\ref{Proof3}) is realized, so the Cauchy-Schwarz inequality tells us that there is a complex number $\lambda$ such that $S_{-}\vert \Phi \rangle = \lambda \vert \Phi \rangle$, so $\vert \Phi \rangle$ is an eigenvector of $S_{-}$ with eigenvalue $\lambda$. From (\ref{Prueba1A}) it follows $\lambda = 0$. Therefore, $S_{-}\vert \Phi \rangle = 0$ and this completes the proof of (\ref{Prueba2A}).

Consider a pure state $\rho_{s} = \vert \Phi \rangle\langle \Phi \vert$. We now prove that
\begin{eqnarray}
\label{Prueba3A}
\mbox{$\rho_{s}$ satisfies (\ref{CondicionSS1})} \Leftrightarrow 
\left\{
\begin{array}{c}
S_{-}\vert \Phi \rangle = (S_{-}^{2})^{\dagger} \vert \Phi \rangle = 0 , \cr
\vert \Phi \rangle \ \ \mbox{is an eigenvector of $S_{11}$.}
\end{array}
\right. \quad\quad
\end{eqnarray} 

Inspecting (\ref{CondicionSS1}) it is clear that (\ref{CondicionSS1}) holds if $S_{-}\vert \Phi \rangle = (S_{-}^{2})^{\dagger} \vert \Phi \rangle = 0$ and $\vert \Phi \rangle$ is an eigenvector of $S_{11}$. 

Now assume that $\rho_{s}$ satisfies (\ref{CondicionSS1}). Since $\mathcal{D}(S_{-})(\rho_{s}) = 0$, it follows from (\ref{Prueba2A}) that $S_{-}\vert \Phi \rangle = 0$. 

Using $S_{-}\vert \Phi \rangle = 0$, the last equation in (\ref{CondicionSS1}) takes the form
\begin{eqnarray}
\label{CondicionSS2}
\alpha_{0}(S_{-}^{2})^{\dagger}\rho_{s} - \alpha_{0}^{*}\rho_{s}S_{-}^{2} = 0 .
\end{eqnarray}
Applying the operator on the lefthand side of (\ref{CondicionSS2}) to $\vert \Phi \rangle$ and using once more that $S_{-}\vert \Phi \rangle = 0$ one obtains that
\begin{eqnarray}
\label{CondicionSS3}
\alpha_{0}(S_{-}^{2})^{\dagger}\vert \Phi \rangle = 0 .
\end{eqnarray}
Since this equation must hold for any complex number $\alpha_{0}$, one concludes that $(S_{-}^{2})^{\dagger}\vert \Phi \rangle = 0$.

Finally, from the first equation in the last line of (\ref{CondicionSS1}) it follows that
\begin{eqnarray}
\label{CondicionSS4}
0 &=& S_{11}\vert \Phi \rangle - \vert \Phi \rangle\langle \Phi \vert S_{11} \vert \Phi \rangle ,
\end{eqnarray}
so $\vert \Phi \rangle$ is an eigenvector of $S_{11}$. This completes the proof of (\ref{Prueba3A}).

At last, we calculate the normalized kets $\vert \Phi \rangle$ that satisfy the conditions that $S_{-}\vert \Phi \rangle = (S_{-}^{2})^{\dagger} \vert \Phi \rangle = 0$ and that $\vert \Phi \rangle$ is an eigenvector of $S_{11}$. 

Assume that 
\begin{eqnarray}
\label{Prueba4A}
S_{-}\vert \Phi \rangle &=& (S_{-}^{2})^{\dagger} \vert \Phi \rangle = 0 , \quad \langle \Phi \vert \Phi \rangle = 1, \cr
S_{11} \vert \Phi \rangle &=& s_{11} \vert \Phi \rangle \quad \mbox{for some real number} \ s_{11}.
\end{eqnarray}

First observe that one can choose $\vert \Phi \rangle$ to have the parity symmetry associated with $\Pi_{0}$ defined in (\ref{pi}). Since $\Pi_{0}$ anticommutes with $S_{-}$ and commutes with both $(S_{-}^{2})^{\dagger}$ and $S_{11}$  [see (\ref{propElogicos})], from (\ref{Prueba4A}) one has
\begin{eqnarray}
\label{sym1}
A\Big[ \Pi_{0} \vert \Phi \rangle \Big] = 0, \quad S_{11} \Big[ \Pi_{0} \vert \Phi \rangle \Big] = s_{11}\Pi_{0} \vert \Phi \rangle ,
\end{eqnarray}
for $A = S_{-}$ and $(S_{-}^{2})^{\dagger}$. Then, $\vert \pm \rangle =  \vert \Phi \rangle \pm \Pi_{0} \vert \Phi \rangle$ satisfy
\begin{eqnarray}
\label{sym2}
S_{-} \vert \pm \rangle  = (S_{-}^{2})^{\dagger} \vert \pm \rangle = 0, \quad S_{11} \vert \pm \rangle = s_{11}\vert \pm \rangle ,
\end{eqnarray}
and at least one of $\vert + \rangle$ and $\vert - \rangle$ is not zero. Therefore, one can look for $\vert \Phi \rangle$ among the kets that are linear combinations of occupation number states $\vert n_{1}, n_{2}, n_{3} \rangle$ where $n_{1}$ is always even or always odd or, equivalently, where $(n_{2} + n_{3})$ is always even or always odd.

Assume $N$ is an even, positive integer. Let
\begin{eqnarray}
\label{phiPM}
\vert \Phi_{+} \rangle &=& \sum_{n_{3} = 0}^{N} \sum_{m = \lceil \frac{n_{3}}{2} \rceil}^{N/2} c(n_{3},m) \vert N-2m, 2m-n_{3}, n_{3} \rangle , \cr
\vert \Phi_{-} \rangle &=& \sum_{n_{3} = 0}^{N-1} \sum_{m = \lceil \frac{n_{3}-1}{2} \rceil}^{\lfloor (N-1)/2 \rfloor} d(n_{3},m) \times \cr
&& \quad\quad\quad \vert N-2m-1, 2m+1-n_{3}, n_{3} \rangle , 
\end{eqnarray}
where $\lceil x \rceil$ and $\lfloor x \rfloor$ are the \textit{ceil} and \textit{floor} functions, that is, $\lceil x \rceil$ is the smallest integer $\geq x$ and $\lfloor x \rfloor$ is the largest integer $\leq x$. Notice that $\vert \Phi_{\pm} \rangle$ have the parity symmetry.

One finds that $S_{-}\vert \Phi_{+} \rangle = 0$ if and only if
\begin{eqnarray}
\label{phiP1}
c(n_{3}, m) &=& c(0,m) (-1)^{n_{3}} \sqrt{ \frac{(2m)!}{(2m-n_{3})! \ n_{3}!} } \ , \quad
\end{eqnarray}
for $m = \lceil n_{3}/2 \rceil ,...,N/2$ and $n_{3} = 0,1,...,N$.
Similarly, $S_{-}\vert \Phi_{-} \rangle = 0$ if and only if
\begin{eqnarray}
\label{phiM1}
d(n_{3}, m) &=& d(0,m) (-1)^{n_{3}} \sqrt{ \frac{(2m+1)!}{(2m+1-n_{3})! \ n_{3}!} } \ , \quad\quad
\end{eqnarray}
for $m = \lceil (n_{3}-1)/2 \rceil ,...,\lfloor (N-1)/2 \rfloor$ and $n_{3} = 0,...,N-1$.

Now, we apply $(S_{-}^{2})^{\dagger}$  to $\vert \Phi_{\pm} \rangle$. Actually, one can choose $\vert \Phi_{+} \rangle$ to be an eigenvector of $S_{-}^{\dagger}$ with eigenvalue $0$. Using (\ref{phiP1}) one obtains that $S_{-}^{\dagger} \vert \Phi_{+} \rangle = 0$ if and only if 
\begin{eqnarray}
\label{phiP2}
c(n_{3},m) &=& 0 \quad \ \mathrm{for} \ \ m= \left\lceil \frac{n_{3}}{2} \right\rceil , ..., \frac{N}{2} -1; \cr
&& \quad\quad\quad \ \  n_{3} = 0, ..., N. 
\end{eqnarray}
Combining (\ref{phiP1}) and (\ref{phiP2}) one concludes that $S_{-}\vert \Phi_{+} \rangle = S_{-}^{\dagger} \vert \Phi_{+} \rangle = 0$ if and only if
\begin{eqnarray}
\label{phiP3}
\vert \Phi_{+} \rangle &=& c\left( 0,\frac{N}{2} \right) \sum_{n_{3}=0}^{N} \vert 0, N-n_{3}, n_{3} \rangle \times \cr
&& \quad\quad\quad (-1)^{n_{3}} \sqrt{\frac{N!}{(N-n_{3})! \ n_{3}! }} . \quad
\end{eqnarray}
Here $c(0,N/2)$ can be used to normalize $\vert \Phi_{+} \rangle$. Choosing $c(0,N/2)>0$ one finds that $\vert \Phi_{+} \rangle$ is normalized if
\begin{eqnarray}
\label{phiP4}
c\left( 0,\frac{N}{2} \right)  &=& 2^{-N/2}.
\end{eqnarray}
In addition, observe that $S_{11}\vert \Phi_{+} \rangle = 0$.

We now consider $\vert \Phi_{-} \rangle$. First, $\vert \Phi_{-} \rangle$ cannot be chosen to be an eigenvector of $S_{-}^{\dagger}$ with eigenvalue zero, so one must use $(S_{-}^{2})^{\dagger}$. Using (\ref{phiM1}) one obtains that $(S_{-}^{2})^{\dagger} \vert \Phi_{-} \rangle = 0$ if and only if 
\begin{eqnarray}
\label{phiM2}
d(n_{3},m) &=& 0 \quad \ \mathrm{for} \ \ m= \left\lfloor \frac{n_{3}}{2} \right\rfloor , ..., \frac{N}{2} -2; \cr
&& \quad\quad\quad \ \ n_{3} = 0,..., N-1. 
\end{eqnarray}
Combining (\ref{phiM1}) and (\ref{phiM2}) one concludes that $S_{-} \vert \Phi_{-} \rangle = (S_{-}^{2})^{\dagger} \vert \Phi_{-} \rangle = 0$ if and only if
\begin{eqnarray}
\label{phiM3}
\vert \Phi_{-} \rangle &=& d\left( 0,\frac{N}{2}-1 \right) \sum_{n_{3}=0}^{N-1} \vert 1, N-1-n_{3}, n_{3} \rangle \times \cr
&& \quad\quad\quad (-1)^{n_{3}} \sqrt{\frac{(N-1)!}{(N-1-n_{3})! \ n_{3}! }} . \quad
\end{eqnarray}
Choosing $d(0,N/2 -1)>0$ one finds that $\vert \Phi_{-} \rangle$ is normalized if
\begin{eqnarray}
\label{phiM4}
d\left( 0,\frac{N}{2} -1 \right)  &=& 2^{-(N-1)/2}.
\end{eqnarray}
In addition, observe that $S_{11}\vert \Phi_{-} \rangle = \vert \Phi_{-} \rangle$.

Using how $b_{j}^{\dagger}$ $(j=1,2,3)$ act on occupation number states, from (\ref{phiP3}), (\ref{phiP4}), (\ref{phiM3}), and (\ref{phiM4}) one has
\begin{eqnarray}
\label{phiPM2}
\vert \Phi_{+} \rangle &=& \frac{1}{\sqrt{2^{N} \ N!}} (b_{2}^{\dagger} - b_{3}^{\dagger})^{N} \vert \mathbf{0} \rangle , \cr
\vert \Phi_{-} \rangle &=& \frac{1}{\sqrt{2^{N-1} \ (N-1)!}} b_{1}^{\dagger}(b_{2}^{\dagger} - b_{3}^{\dagger})^{N-1} \vert \mathbf{0} \rangle ,
\end{eqnarray}
where $\vert \mathbf{0} \rangle = \vert n_{1}=0, n_{2} = 0, n_{3} = 0 \rangle$. Using the definition of SU(3) coherent states in (\ref{coherent1}), from (\ref{phiPM2}) one concludes that 
\begin{eqnarray}
\label{phiPM3}
\vert \Phi_{+} \rangle &=& \vert z_{1} = 0, z_{2} = 1, z_{3} = -1 \rangle_{N} , \cr
\vert \Phi_{-} \rangle &=& b_{1}^{\dagger} \vert z_{1} = 0, z_{2} = 1, z_{3} = -1 \rangle_{N-1}.
\end{eqnarray}
Therefore, when $N$ is an even, positive integer one finds that (\ref{Prueba4A}) holds if and only if $\vert \Phi \rangle$ is equal (except for a global phase factor) to $\vert \Phi_{+} \rangle$ or $\vert \Phi_{-} \rangle$.

Now assume that $N$ is an odd, positive integer. Motivated by the case where $N$ is an even, positive integer, propose
\begin{eqnarray}
\vert \Phi_{+} \rangle &=& b_{1}^{\dagger} \vert z_{1} = 0, z_{2}=1, z_{3} = -1 \rangle_{N-1} , \cr
\vert \Phi_{-} \rangle &=&  \vert z_{1} = 0, z_{2}=1, z_{3} = -1 \rangle_{N} .
\end{eqnarray}
Then, it is straightforward to verify that
\begin{eqnarray}
\label{phiPM4}
S_{-} \vert \Phi_{\pm} \rangle &=& 0 , \quad (S_{-}^{2})^{\dagger} \vert \Phi_{\pm} \rangle = 0, \quad S_{11} \vert \Phi_{+} \rangle = \vert \Phi_{+} \rangle , \cr
S_{-}^{\dagger} \vert \Phi_{+} \rangle &\not=& 0 , \quad  \ S_{-}^{\dagger} \vert \Phi_{-} \rangle = 0 , \quad\quad S_{11} \vert \Phi_{-} \rangle = 0 . \quad\quad\quad
\end{eqnarray}


\section{Averaging}
\label{AppendixC}

We present \textit{James' averaging method} \cite{James1,James2} to third order (the reference only presents it to second order). It allows one to obtain an effective Hamiltonian that accounts for the most important effects of an interaction when considering a rapidly oscillating Hamiltonian.

Consider a quantum system whose evolution is governed by a Hamiltonian $H(t)$. Note that $H(t)$ may or may not be explicitly time-dependent. We denote the state of the system at time $t$ by $\vert \psi (t) \rangle$.

Now pass to an interaction picture (IP) defined by the unitary operator
\begin{eqnarray}
\label{James1}
U_{I}(t,t_{0}) &=& e^{-\frac{i}{\hbar}H_{0}(t-t_{0})} ,
\end{eqnarray}
where $H_{0}$ is a time-independent Hermitian operator and $t_{0}$ is a fixed time. 

The state of the system in the IP is given by
\begin{eqnarray}
\label{James2}
\vert \psi_{I}(t) \rangle &=& U_{I}^{\dagger} (t,t_{0}) \vert \psi (t) \rangle , 
\end{eqnarray}
and the IP Schr\"{o}dinger equation is
\begin{eqnarray}
\label{James3}
i\hbar \frac{d}{dt} \vert \psi_{I}(t) \rangle &=& H_{I}(t) \vert \psi_{I} (t) \rangle,
\end{eqnarray}
with 
\begin{eqnarray}
\label{James4}
H_{I}(t) &=& U_{I}^{\dagger} (t,t_{0}) H(t) U_{I}(t,t_{0}) - H_{0} .
\end{eqnarray}
Then, the IP evolution operator $U_{IP}(t,t_{0})$ is defined by the initial value problem
\begin{eqnarray}
\label{James5}
i\hbar \frac{\partial}{\partial t} U_{IP} (t,t_{0}) &=& H_{I}(t) U_{IP}(t,t_{0}), \quad U_{IP}(t_{0},t_{0}) = \mathbb{I},  \quad\quad
\end{eqnarray} 
where $\mathbb{I}$ is the identity operator.

We now introduce the averaging. Given a (possibly time-dependent) linear operator $A(t)$, define the averaged linear operator $\overline{A(t)}$ by
\begin{eqnarray}
\label{James6}
\overline{A(t)} &=& \int_{-\infty}^{+\infty}dt' f(t-t')A(t') ,
\end{eqnarray}
where $f(t)$ is a real-valued function such that $\int_{-\infty}^{+ \infty} dt \ f(t) = 1$. In principle, one should also demand that $f(t) = 0$ for $t <0$ to avoid problems with causality but this condition can be omitted in the present context. The purpose of the function $f(t)$ is to eliminate high-frequency terms by averaging them to zero. We use an \textit{ideal low pass filter}
\begin{eqnarray}
\label{James7}
f(t) &=& \frac{\mathrm{sin}(\omega_{0}t)}{\pi t} , 
\end{eqnarray}
whose Fourier transform is
\begin{eqnarray}
\label{James8}
\hat{f}(\omega) &=& \frac{1}{\sqrt{2 \pi}} \int_{-\infty}^{+\infty} dt \ f(t)e^{-i\omega t} , \cr
&=& \frac{\theta (\omega + \omega_{0})\theta(\omega_{0}- \omega)}{\sqrt{2\pi}} .
\end{eqnarray}
Here $\omega_{0} >0$ is a cutoff frequency and $\theta$ is the Heaviside step-function: $\theta (x) = 1$ if $x \geq 0$ and $\theta (x) = 0$ if $x <0$.
To illustrate how the filter works consider a linear operator of the form $A(t) = A_{0}e^{-i \Omega (t-t_{0})}$ where $\Omega$ is a real quantity. Using (\ref{James6})-(\ref{James8}), the strong Parseval formula, and the Dirac delta function $\delta(\omega - \Omega)$ one has
\begin{eqnarray}
\label{James6b}
\overline{A(t)} &=& A(t) \int_{-\infty}^{+\infty}d\tau \ f(\tau)e^{i\Omega \tau} , \cr
&=&  A(t) \int_{-\infty}^{+\infty} d\omega \hat{f}(\omega) \sqrt{2\pi} \delta (\omega-\Omega), \cr
&=& \left\{
\begin{array}{cc}
A(t) & \mathrm{if} \ \ \Omega \in (-\omega_{0}, \omega_{0}), \cr
0 & \mbox{in any other case.}
\end{array}
\right.
\end{eqnarray}
Hence, the averaging eliminates high frequency terms with $\vert \Omega \vert > \omega_{0}$. 

From (\ref{James6}) it is straightforward to show that 
\begin{eqnarray}
\label{James9}
\overline{A^{\dagger}(t)} &=& \Big[ \overline{A(t)} \Big]^{\dagger} , \quad \frac{d}{dt} \overline{A(t)} = \overline{\frac{dA}{dt}(t)} .
\end{eqnarray}

Averaging (\ref{James5}) and using (\ref{James9}) one obtains 
\begin{eqnarray}
\label{James10}
i\hbar \frac{\partial}{\partial t} \overline{U_{IP} (t,t_{0})} &=& \overline{H_{I}(t) U_{IP}(t,t_{0})} .  \quad\quad \ \
\end{eqnarray}
Motivated by (\ref{James10}), define the operator $\mathsf{H}_{\mathrm{avg}}(t)$ by
\begin{eqnarray}
\label{James11}
i\hbar \frac{\partial}{\partial t} \overline{U_{IP} (t,t_{0})} &=& \mathsf{H}_{\mathrm{avg}}(t) \overline{U_{IP}(t,t_{0})} .  \quad\quad \ \
\end{eqnarray}
From (\ref{James10}) and (\ref{James11}) it follows that
\begin{eqnarray}
\label{James12}
\mathsf{H}_{\mathrm{avg}}(t) &=& \overline{H_{I}(t) U_{IP}(t,t_{0})} \Big[  \overline{U_{IP}(t,t_{0})} \Big]^{-1}.
\end{eqnarray}
Usually $\mathsf{H}_{\mathrm{avg}}(t)$ in (\ref{James12}) is not Hermitian. The averaged Hamiltonian is then defined to be
\begin{eqnarray}
\label{James13}
H_{\mathrm{avg}}(t) &=& \frac{1}{2} \left[ \mathsf{H}_{\mathrm{avg}}(t) + \mathsf{H}_{\mathrm{avg}}^{\dagger} (t)\right] .
\end{eqnarray} 

Now assume that $H_{I}(t)$ is a perturbation. Then one has the expansion
\begin{eqnarray}
\label{James14}
U_{IP}(t,t_{0}) &=& \mathbb{I} + \sum_{n=1}^{+\infty} U_{IP}^{(n)} (t,t_{0}) ,
\end{eqnarray}
where
\begin{eqnarray}
\label{James15}
U_{IP}^{(1)}(t,t_{0}) &=& -\frac{i}{\hbar} \int_{t_{0}}^{t} dt_{1} H_{I}(t_{1}) , \cr
U_{IP}^{(n)}(t,t_{0}) &=& \left( -\frac{i}{\hbar} \right)^{n} \int_{t_{0}}^{t} dt_{1} \int_{t_{0}}^{t_{1}}dt_{2} ... \int_{t_{0}}^{t_{n-1}} dt_{n} H_{I}(t_{1}) \times \cr
&& \quad H_{I}(t_{2}) ... H_{I}(t_{n}) , \quad\quad (n=2,3,...).
\end{eqnarray}
Notice that 
\begin{eqnarray}
\label{James15b}
\Big[ U_{IP}^{(1)}(t,t_{0}) \Big]^{\dagger} &=& - U_{IP}^{(1)}(t,t_{0}).
\end{eqnarray}

Using the linearity of the average (\ref{James6}), the expansion in (\ref{James14})-(\ref{James15b}), and the Maclaurin series of $1/(1+x)$ it is straightforward to show that
\begin{eqnarray}
\label{James16}
\overline{H_{I}(t) U_{IP}(t,t_{0})} 
&=& \overline{H_{I}(t)} + \sum_{n=1}^{+\infty} \overline{H_{I}(t) U_{IP}^{(n)}(t,t_{0})} , \quad\quad
\end{eqnarray}
and
\begin{eqnarray}
\label{James17}
&& \Big[\overline{U_{IP}(t,t_{0})} \Big]^{-1}
= 
\Bigg[ \mathbb{I} + \sum_{n=1}^{+\infty} \overline{U_{IP}^{(n)} (t,t_{0})} \Bigg]^{-1} , \cr
&=& \mathbb{I} - \overline{U_{IP}^{(1)} (t,t_{0})} - \overline{U_{IP}^{(2)} (t,t_{0})} - \overline{U_{IP}^{(3)} (t,t_{0})} \cr
&& + \Bigg[ \overline{U_{IP}^{(1)} (t,t_{0})} \Bigg]^{2} + \Bigg[ \overline{U_{IP}^{(1)} (t,t_{0})} \Bigg] \Bigg[ \overline{U_{IP}^{(2)} (t,t_{0})} \Bigg] \cr
&& + \Bigg[ \overline{U_{IP}^{(2)} (t,t_{0})} \Bigg] \Bigg[ \overline{U_{IP}^{(1)} (t,t_{0})} \Bigg] - \Bigg[ \overline{U_{IP}^{(1)} (t,t_{0})} \Bigg]^{3} + ... \ . \quad\quad \cr
&&
\end{eqnarray}
Here and in the following the dots $...$ indicate terms of order $\geq 4$ in $H_{I}(t)$. Then, it follows from (\ref{James12}), (\ref{James16}), and (\ref{James17}) that
\begin{eqnarray}
\label{James18}
\mathsf{H}_{\mathrm{avg}} (t)
&=& \overline{H_{I}(t)} \Bigg\{ \mathbb{I} - \overline{U_{IP}^{(1)} (t,t_{0})} - \overline{U_{IP}^{(2)} (t,t_{0})}  \cr
&&  + \Bigg[ \overline{U_{IP}^{(1)} (t,t_{0})} \Bigg]^{2} \Bigg\} + \overline{H_{I}(t)U_{IP}^{(1)} (t,t_{0})} \cr
&& - \Bigg[ \overline{H_{I}(t)U_{IP}^{(1)} (t,t_{0})} \Bigg] \overline{U_{IP}^{(1)} (t,t_{0})} \cr
&& + \overline{H_{I}(t)U_{IP}^{(2)} (t,t_{0})} + ... \ .
\end{eqnarray}

Observe from (\ref{James18}) that it is advantageous to choose (when possible) $H_{0}$ such that
\begin{eqnarray}
\label{James19}
\overline{H_{I}(t)} &=& 0 ,
\end{eqnarray}
because all terms between the curly brackets disappear. In particular, all the terms of order $1$ in $H_{I}(t)$ are eliminated. In the following we assume that (\ref{James19}) holds.

Neglecting terms of order $\geq 4$ in $H_{I}(t)$, it follows from (\ref{James13}), (\ref{James18}), and (\ref{James19}) that
\begin{eqnarray}
\label{James20}
H_{\mathrm{avg}} (t)
&\simeq& H_{\mathrm{avg}}^{(2)} (t) + H_{\mathrm{avg}}^{(3)} (t),
\end{eqnarray}
where $H_{\mathrm{avg}}^{(j)} (t)$ includes the terms of order $j$ $(j=2,3)$ in $H_{I}(t)$ and is given by
\begin{eqnarray}
\label{James21} 
H_{\mathrm{avg}}^{(2)} (t) &=& \frac{1}{2} \overline{ [ H_{I}(t), \ U_{IP}^{(1)}(t,t_{0}) ] } , \cr
H_{\mathrm{avg}}^{(3)} (t) &=& \frac{1}{2} \Bigg\{ \ \overline{H_{I}(t)U_{IP}^{(2)}(t,t_{0})}  + \Bigg[ \overline{H_{I}(t)U_{IP}^{(2)}(t,t_{0})} \Bigg]^{\dagger} \cr
&& \quad\quad - \Bigg[ \overline{H_{I}(t) U_{IP}^{(1)}(t,t_{0})} \Bigg] \ \overline{U_{IP}^{(1)}(t,t_{0})} \cr
&& \quad\quad - \overline{U_{IP}^{(1)}(t,t_{0})} \ \Bigg[ \overline{U_{IP}^{(1)}(t,t_{0}) H_{I}(t)} \Bigg] \  \Bigg\} . \quad\quad
\end{eqnarray}

We now apply the averaging method with the same notation as that used above. Consider the Hamiltonian 
\begin{eqnarray}
\label{JamesP1}
H &=& H_{0} + H_{\mathrm{int}}
\end{eqnarray}
where $H_{\mathrm{int}}$ is defined in (\ref{Origen11}) and
\begin{eqnarray}
\label{James22}
\frac{1}{\hbar} H_{0} &=& \delta_{p} a_{p}^{\dagger}a_{p} + \delta_{s} a_{s}^{\dagger}a_{s} + \sum_{j=2,3} \omega_{j} S_{jj} +  \frac{\omega_{d}}{2} S_{11} . \quad\quad 
\end{eqnarray}
Observe that $H$ in (\ref{JamesP1}) coincides with $H_{I}$ in (\ref{Origen6}) when one takes $\Omega_{d} = 0$. We take $\Omega_{d} = 0$ because we want to obtain an approximation for the \textit{qutrits-signal} and \textit{pump-signal} interactions described by $H_{\mathrm{int}}$. 

First pass to the IP by means of the unitary transformation in (\ref{James1}) with $H_{0}$ in (\ref{James22}). Then, the IP Schr\"{o}dinger equation (\ref{James3}) has the Hamiltonian
\begin{eqnarray}
\label{James24}
\frac{1}{\hbar}H_{I}(t) &=& \frac{1}{\hbar} U_{I}^{\dagger}(t) H_{\mathrm{int}} U_{I}(t) , \cr
&=& J \left[ a_{p} (a_{s}^{\dagger})^{2} e^{i \zeta_{1} t} + a_{p}^{\dagger} a_{s}^{2} e^{-i\zeta_{1}t} \right] \cr
&& + a_{s}^{\dagger}\sum_{j=2,3} g_{j} S_{1j} e^{i \zeta_{j} t} + a_{s}\sum_{j=2,3} g_{j} S_{1j}^{\dagger} e^{-i \zeta_{j} t}  , \cr
&=& \frac{1}{\hbar}\sum_{j=1,2,3} \left( F_{j}e^{-i\zeta_{j}t} + F_{j}^{\dagger}e^{i\zeta_{j}t} \right).
\end{eqnarray}
Here we have introduced the operators and the detunings
\begin{eqnarray}
\label{James25}
F_{1} &=& \hbar J a_{p}^{\dagger}a_{s}^{2} , \quad \ \ F_{j} = \hbar g_{j} a_{s}S_{1j}^{\dagger},  \cr
\zeta_{1} &=& 2\omega_{s} -\omega_{p} , \quad \zeta_{j} = \omega_{s} - \omega_{j} , \quad (j=2,3) . \quad
\end{eqnarray}
The form of $H_{I}(t)$ in the last line of (\ref{James24}) is convenient to carry out the lengthy calculations of the method. 

In the following assume that (\ref{CondicionAvg}) and (\ref{James25b}) hold. Then, $\zeta_{1} \geq \omega_{s}$ and $\zeta_{j} \sim \omega_{s}$ for $j=2,3$.

Consider a cutoff frequency $\omega_{0}>0$ such that
\begin{eqnarray}
\label{James26}
&& \vert \zeta_{j} + \zeta_{k} - \zeta_{1} \vert = \vert \omega_{j} + \omega_{k} - \omega_{p} \vert < \omega_{0}  \quad (j,k=2,3), \cr
&& \vert \zeta_{3} - \zeta_{2} \vert = \vert \omega_{3} - \omega_{2} \vert < \omega_{0} ,
\end{eqnarray} 
and 
\begin{eqnarray}
\label{James27}
n\zeta_{l_{1}}, \ \zeta_{2} + \zeta_{3}, \ \zeta_{1} \pm \zeta_{j}, \zeta_{l_{1}} + \zeta_{l_{2}} + \zeta_{l_{3}}  &>& \omega_{0} , \cr
\vert \zeta_{l_{1}} + \zeta_{l_{2}} - \zeta_{l_{3}} \vert &>& \omega_{0} , \quad\quad
\end{eqnarray}
for $l_{1},l_{2},l_{3}=1,2,3$ and $n=1,2$ and $j,k=2,3$ with $\vert \zeta_{l_{1}} + \zeta_{l_{2}} - \zeta_{l_{3}} \vert$ not of the form $\vert \zeta_{j} + \zeta_{k} - \zeta_{1} \vert$. The assumption in (\ref{James25b}) guarantees that the quantities that appear on the lefthand side of the inequalities in (\ref{James27}) are $\gtrsim \omega_{s} $, while the quantities that appear on the lefthand side of the inequalities in (\ref{James26}) are $\ll \omega_{s}$. 

It is important to note that (\ref{CondicionAvg}) must hold in order to consider $H_{I}(t)$ a perturbation. The reason for this is that, when one expresses the IP Schr\"{o}dinger equation in terms of the dimensionless time $\tau = \omega_{s}t$, all the parameters $J$, $g_{2}$, and $g_{3}$ in $H_{I}(t)$ are divided by $\omega_{s}$ and one requires the quotients $J/\omega_{s}$, $g_{2}/\omega_{s}$, and $g_{3}/\omega_{s}$ to be small so that high frequency terms average to zero. 

Applying the averaging method with $t_{0}=0$ and with (\ref{James26}) and (\ref{James27}) leads to $\overline{H_{I}(t)} = 0$ and, consequently, to the following approximation to $3^{\mathrm{rd}}$ order in $H_{I}(t)$: 
\begin{eqnarray}
\label{James28}
H_{I}(t)
&\simeq& H_{\mathrm{avg}}^{(2)}(t) + H_{\mathrm{avg}}^{(3)}(t) ,
\end{eqnarray} 
where $H_{\mathrm{avg}}^{(2)}(t)$ and $H_{\mathrm{avg}}^{(3)}(t)$ are given by (\ref{James21}). 

Returning to the original picture, one obtains from (\ref{James24}) and (\ref{James28}) that
\begin{eqnarray}
\label{James31}
H_{\mathrm{int}} &=& U_{I}(t) H_{I}(t) U_{I}^{\dagger}(t) , \cr
&\simeq& U_{I}(t) \left[ H_{\mathrm{avg}}^{(2)}(t) + H_{\mathrm{avg}}^{(3)}(t) \right] U_{I}^{\dagger}(t).
\end{eqnarray}
Carrying out the calculations and taking the expected value in the state $\vert 0_{s} \rangle$ leads to (\ref{Averaging}).

\section{Approximate evolution}
\label{AppF}

In this appendix we show that (\ref{ExtraF5}) holds if (\ref{Final}) and the assumptions of the first (\ref{Origen8}) and second (\ref{Origen16}) adiabatic approximations are satisfied [see also the paragraphs below (\ref{Origen8}) and (\ref{Origen16})]. 

To have succinct expressions we introduce the density operators of the pump+qutrits subsystem
\begin{eqnarray}
\label{AppF1}
\rho_{pq}^{(\mathbb{j})} &=&  \vert \alpha_{po} \rangle\langle \alpha_{po} \vert \otimes \vert \mathbb{j}_{L} \rangle\langle \mathbb{j}_{L} \vert \quad\quad (\mathbb{j}=0,1) ,
\end{eqnarray}
where $\vert 0_{L} \rangle$ and $\vert 1_{L} \rangle$ are defined in (\ref{edoEstacionario1}) and $\vert \alpha_{po} \rangle$ is a coherent state with $\alpha_{po}$ in (\ref{ExtraF3}).

In all that follows it is used without mention that $\vert 0_{L} \rangle$ and $\vert 1_{L} \rangle$ satisfy the properties in (\ref{propElogicos}), that $\vert 0_{s}\rangle\langle 0_{s} \vert$ is the steady-state solution of the damped harmonic oscillator master equation at zero temperature in (\ref{Origen9}), and that $\vert \alpha_{po} \rangle\langle \alpha_{po} \vert$ is the steady-state solution of the driven and damped harmonic oscillator master equation at zero temperature in (\ref{Origen17}) with $\delta_{p}$ and $\kappa_{p}$ replacing $\delta_{p}''$ and $\kappa_{p}'$, respectively.

First, it is straightforward to show from (\ref{Origen5}), (\ref{ExtraF1}), and (\ref{ExtraF2}) that one has
\begin{eqnarray}
\label{ExtraF4}
\mathcal{L}_{I}\rho_{\scriptscriptstyle{\mathbb{j}}} 
&=& -i\sqrt{2} J \Big[ \alpha_{po} \vert 2_{s} \rangle\langle 0_{s} \vert - \alpha_{po}^{*} \vert 0_{s} \rangle\langle 2_{s} \vert \Big] \otimes \rho_{pq}^{(\mathbb{j})} , \quad\quad
\end{eqnarray}
and that
\begin{eqnarray}
\label{AppF2}
\mathcal{L}_{I} \vert 0_{s} \rangle\langle 2_{s} \vert \otimes \rho_{pq}^{(0)} &=& \Bigg[ (-\kappa_{s} + i2\delta_{s})\vert 0_{s} \rangle\langle 2_{s} \vert + R_{s}^{(0)} \Bigg] \otimes \rho_{pq}^{(0)} \cr
&& + i \sqrt{2} J \vert 0_{s}\rangle\langle 0_{s} \vert \otimes \rho_{pq}^{(0)}a_{p} , \cr
\mathcal{L}_{I} \vert 0_{s} \rangle\langle 2_{s} \vert \otimes \rho_{pq}^{(1)} &=& \Bigg[ (-\kappa_{s} + i2\delta_{s})\vert 0_{s} \rangle\langle 2_{s} \vert + R_{s}^{(0)} \Bigg] \otimes \rho_{pq}^{(1)} \cr
&& + i \sqrt{2} J \vert 0_{s}\rangle\langle 0_{s} \vert \otimes \rho_{pq}^{(1)}a_{p} \cr 
&& +i\sqrt{2} g\vert 0_{s} \rangle\langle 1_{s} \vert \otimes \rho_{pq}^{(1)}S_{-} ,
\end{eqnarray}
where
\begin{eqnarray}
\label{AppF3}
R_{s}^{(0)} &=& -i\sqrt{2} J \alpha_{po} \vert 2_{s} \rangle\langle 2_{s} \vert + i \sqrt{12} J \alpha_{po}^{*} \vert 0_{s} \rangle\langle 4_{s} \vert  . \quad
\end{eqnarray}

Now lets compare the magnitude of the coefficients of the terms appearing in (\ref{AppF2}) and (\ref{AppF3}):
\begin{eqnarray}
\label{AppF4}
\left\vert \frac{i \sqrt{2} J}{-\kappa_{s} + i 2\delta_{s}} \right\vert &\leq& \sqrt{2} \epsilon_{1}  ,\quad \ \ \ \left\vert \frac{i\sqrt{2} g}{-\kappa_{s} + i 2\delta_{s}}  \right\vert \leq \sqrt{2} \epsilon_{1}, \cr
\left\vert \frac{-i \sqrt{2} J \alpha_{po}}{-\kappa_{s} + i 2\delta_{s}} \right\vert &\lesssim& 2\sqrt{2} \epsilon_{1} , \quad \ \ \left\vert \frac{i \sqrt{12} J \alpha_{po}^{*}}{-\kappa_{s} + i 2\delta_{s}} \right\vert \lesssim 2\sqrt{12} \epsilon_{1} , \quad\quad
\end{eqnarray}
where $\epsilon_{1}$ is defined in (\ref{Origen8}) and where $\lesssim$ appears instead of $\leq$ because the second adiabatic approximation requires $\vert \Omega_{d}/\kappa_{p} \vert \sim 1$ [see the paragraph below (\ref{Origen16})]. Since the first adiabatic approximation requires $\epsilon_1 \ll 1$, to good approximation one can neglect all the terms appearing on the righthand side of (\ref{AppF2}) except for the first one:
\begin{eqnarray}
\label{AppF5}
\mathcal{L}_{I} \vert 0_{s} \rangle\langle 2_{s} \vert \otimes \rho_{pq}^{(\mathbb{j})} &\simeq& (-\kappa_{s} + i2\delta_{s})\vert 0_{s} \rangle\langle 2_{s} \vert \otimes \rho_{pq}^{(\mathbb{j})} . \quad
\end{eqnarray}
Since 
\begin{eqnarray}
\label{AppF6}
\mathcal{L}_{I} \vert 2_{s} \rangle\langle 0_{s} \vert \otimes \rho_{pq}^{(\mathbb{j})}  &=& \left[ \mathcal{L}_{I} \vert 0_{s} \rangle\langle 2_{s} \vert \otimes \rho_{pq}^{(\mathbb{j})}  \right]^{\dagger},
\end{eqnarray} 
from (\ref{ExtraF4}) and (\ref{AppF5}) one can obtain (\ref{ExtraF5}).


\bibliography{cites}

\end{document}